# Manufacturing Cybersecurity from Threat to Action: A Taxonomy-Guided Decision Support Framework

**Md Habibor Rahman[a], Rocco Cassandro[b], Thorsten Wuest[c], Mohammed Shafae[d, 1]**

[a] Department of Mechanical Engineering, University of Massachusetts Dartmouth, North Dartmouth, MA 02747, USA
[b] Department of Industrial Engineering and Engineering Management, Western New England University, Springfield, MA 01119, USA
[c] Department of Mechanical Engineering, University of South Carolina, Columbia, SC 29208, USA
[d] Department of Systems and Industrial Engineering, The University of Arizona, Tucson, AZ 85721, USA

## Abstract

An attack taxonomy is essential for defending manufacturing systems against cyber-physical threats by enabling systematic understanding and classification of threat attributes. However, existing taxonomies typically focus on a limited set of attributes and fail to comprehensively integrate threat actors, system-level and operational impacts, and potential countermeasures within a unified framework. Additionally, converting taxonomy-based knowledge into actionable guidance for cybersecurity tool development and decision-making remains challenging and understudied. To address these gaps, this study introduces a comprehensive attack-countermeasure taxonomy along with a taxonomy-guided decision-support framework, providing an end-to-end approach from threat identification to mitigation in manufacturing systems. Specifically, the proposed taxonomy classifies threat actors and their intent, system behavioral deviations during threat events, attack methods, and attack targets and incorporates both operational and system-level impacts. Furthermore, a structured classification of countermeasures is integrated within the taxonomy, supported by illustrative examples of potential countermeasures. Unlike previous taxonomies, this model captures the entire attack chain—from adversarial intent to observable system deviations and corresponding countermeasures. The taxonomy's practical implementation is demonstrated using realistic attack scenarios, real-world incidents, and relevant academic case studies. Building upon this foundation, the proposed taxonomy-guided decision-support framework shows explicitly how each taxonomy layer helps guide threat identification, risk modeling and assessment, and appropriate countermeasure selection and deployment. Moreover, the framework highlights how the taxonomy complements existing cybersecurity tools, frameworks, and methodologies to facilitate context-aware and risk-informed security decisions in smart manufacturing environments.

## 1 Introduction

The convergence of Information Technology (IT) and Operational Technology (OT) has transformed previously air-gapped and isolated manufacturing systems into data-driven, distributed, and interconnected Smart Manufacturing Systems (SMS). While digital manufacturing technologies have enabled adaptive decision-making and control, increased production automation, and real-time data-driven operations, their integration with physical manufacturing assets (e.g., machine tools) poses significant cybersecurity threats to SMS (Rahman et al., 2024). The proliferation of cybersecurity risk in manufacturing systems is outpacing most organizations' ability to respond, widening the cyber risk gap (*Deloitte*, 2022). In 2024, manufacturing accounted for 26% of all cyberattacks among the top ten industries, marking its fourth consecutive year as the most targeted sector (*IBM*, 2025). In response, defending SMS against the growing threat of cyberattacks must be a key priority in manufacturing digitalization efforts. An essential step towards a more proactive approach is characterizing cybersecurity threat attributes and countermeasures thoroughly and systematically in the context of SMS. This can also help to mitigate the consequences of an attack and ensure that the manufacturing system can restore regular operations as soon as possible after encountering threat events.

[1] Corresponding Author.
*E-mail address:* shafae1@arizona.edu (M. Shafae). ORCID: 0000-0002-4060-906X.

The critical threat attributes and their relationship with the countermeasure development process are depicted in Fig. 1. Cybersecurity threat events encompass adversaries, their motives, and attack methods to attack impacts. The *threat actor* represents adversaries with the potential to harm a system, the *attack method* represents how threat actors can compromise the system, *attack targets* indicate where threat actors can infiltrate the digital manufacturing value chain, and *attack impacts* are the results of an attack on system assets (Rahman, 2024). A systematic classification and characterization of all key elements of cyber-physical attacks on SMS enables stakeholders to compare and contrast different cyberattacks, identify patterns and trends in threat events, assess the associated cybersecurity risk, and develop strategies to defend against them. In pursuit of this research objective, a cyberattack taxonomy is an essential tool for improving cybersecurity for three key reasons. *First*, taxonomies provide a uniform and structured classification framework to understand, identify, and categorize threat attributes, such as threat actors, attack methods, targets, impacts, and available countermeasures. *Second*, taxonomies enable manufacturing firms to better understand the threats to which they may be exposed and take the necessary precautions to protect themselves. *Third*, taxonomies enable the research community to easily share knowledge between researchers and practitioners through a consistent classification scheme.

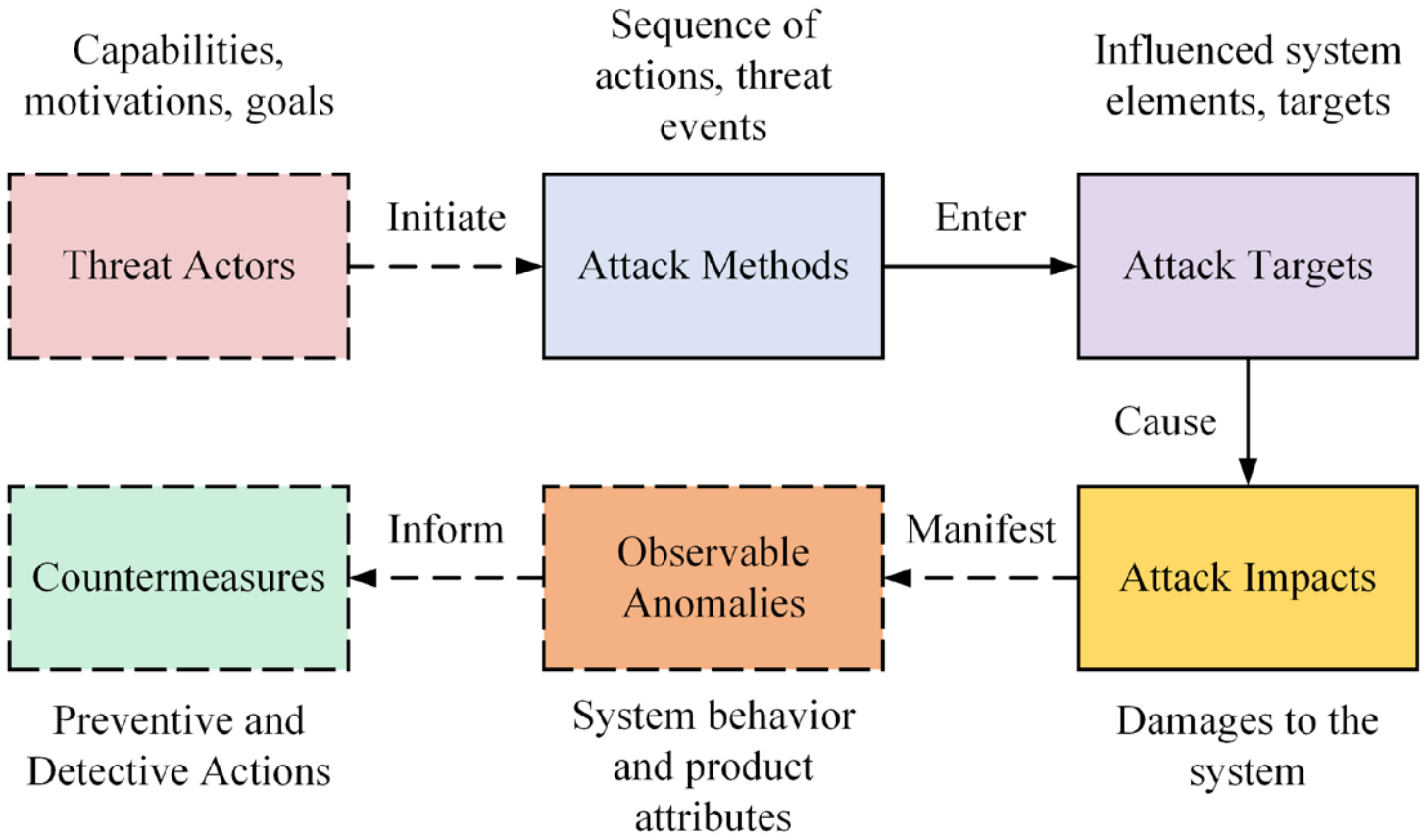

**Fig. 1**: A holistic view of the manufacturing-specific cybersecurity threat attributes and countermeasures in smart manufacturing systems (dashed line denotes the missing elements in current manufacturing cyberattack taxonomies)

Acknowledging the significance, several taxonomies and classification schemes have been proposed to provide taxonomical classifications of manufacturing-specific cybersecurity threat attributes to develop a common language for understanding potential attacks and countermeasures. An overview of manufacturing cyberattack taxonomies is provided in Table 1, summarizing the significant contribution(s) and the specific attack attributes covered taxonomically in current taxonomies. Similar to the Common Attack Pattern Enumeration and Classification (CAPEC[2]) (*CAPEC*, 2023), a community resource for identifying and understanding attacks, most taxonomies distinguished attacks by primarily focusing on attack targets[3] and attack methods (Elhabashy et al., 2019; Mahesh et al., 2021; Pan et al., 2017; Shafae et al., 2019; D. Wu et al., 2018; M. Wu & Moon, 2017, 2018; Yampolskiy et al., 2018). Researchers also categorized attack impact on SMS as a critical attribute and included them in the attack description (Elhabashy et al., 2019; Pan et al., 2017; Shafae et al., 2019; Williams et al., 2023; D. Wu et al., 2018; M. Wu & Moon, 2017, 2018; Yampolskiy et al., 2018). While most taxonomies discussed attack impacts such as theft of confidential information and product-level consequences (e.g., degraded quality), there is little to no focus on system-level and operational implications of cyberattacks in production processes (Rahman et al., 2023). A few studies present general overviews of countermeasures, but they do so without structured classification or integration into a taxonomical framework.

---

[2] CAPEC is sponsored by the U.S. Department of Homeland Security (DHS) Cybersecurity and Infrastructure Security Agency (CISA) and managed by the Homeland Security Systems Engineering and Development Institute (HSSEDI) which is operated by The MITRE Corporation (MITRE) (*CAPEC*, 2023).

[3] While attack targets and attack locations are interchangeably used in manufacturing cyberattack taxonomies, this paper is following the NIST definition from (Framework for Improving Critical Infrastructure Cybersecurity, Version 1.1, 2018)

**Table 1:** Recent literature on manufacturing cyberattack taxonomy

| References | Major Contributions | Taxonomical classification provided for | | | | | |
|---|---|---|---|---|---|---|---|
| | | Threat Actors | Attack methods | Attack targets | Attack impacts | Observable anomalies | Counter measures |
| (Pan et al., 2017) | Classified cyber-physical attacks on IoT-based manufacturing processes based on four elements of the attack flow. | | ☑ | ☑ | ☑ | | |
| (M. Wu & Moon, 2017) | Proposed taxonomy for cross-domain attacks with four dimensions, focusing on their cyber and physical aspects | | ☑ | ☑ | ☑ | | |
| (Tuptuk & Hailes, 2018) | Presented an adversarial model; categorized adversaries and manufacturing-specific cyberattack types | ☑ | ☑ | | | | |
| (D. Wu et al., 2018) | Briefly reviewed the cybersecurity threat in SMS and classified countermeasures. | | ☑ | ☑ | ☑ | | ☑ |
| (Yampolskiy et al., 2018) | Provided in-depth classification of attack targets and methods for technical data theft and AM sabotage. | | ☑ | ☑ | | | |
| (M. Wu & Moon, 2018) | Further categorized attack targets into sub-targets and provided detailed cyber-physical attack methods. | | ☑ | ☑ | ☑ | | |
| (Elhabashy et al., 2019) | Integrated quality control with cyber-physical attacks and proposed an attack taxonomy with four layers. | | ☑ | ☑ | | | |
| (Shafae et al., 2019) | Presented an attack design scheme for manufacturing cyberattacks and classified attack design considerations | | | ☑ | ☑ | | |
| (Mahesh et al., 2021) | Presented cybersecurity risks in CPMS, their impacts, and defense approaches to secure manufacturing. | | ☑ | ☑ | | | ☑ |
| (Williams et al., 2023) | Surveyed cyberattacks in the manufacturing industry and categorized them according to their impacts. | | | | ☑ | | |
| This paper | Proposes an attack taxonomy for a holistic characterization of cybersecurity threat attributes and countermeasures. | ☑ | ☑ | ☑ | ☑ | ☑ | ☑ |

Current work often focuses on a narrow range of attacks and omits critical attributes of manufacturing attacks, such as threat actors and their intentions, specific system/machine behavioral deviations, and system damages resulting from cyberattacks. Understanding threat actors systematically is essential for comprehending the security risk, adversaries' capabilities, prospective attack motives, and strategies, all of which are required for defending against them (Blank & Gallagher, 2012; Sailio et al., 2020). Anomalies in system behavior or product attributes can aid in designing and developing attack prevention and detection techniques because these anomalies manifest attacks on physical targets. Current taxonomies provide insights into attack propagation but lack attack realization through observed system deviations. Additionally, while cyberattack taxonomies are valuable in enabling a comprehensive and consistent understanding of threat attributes, it is unclear whether and how such structured knowledge can be incorporated into the development of cybersecurity tools in applications such as risk assessment and mitigation and what the implications of the taxonomy are to practitioners (Rahman et al., 2023). This gap limits the practical utility of taxonomies in real-world manufacturing environments. There is also insufficient guidance on how taxonomical frameworks can be coupled with other security tools, frameworks, and methodologies to support context-aware and risk-informed decision-making.

In response to these challenges, this study aims to develop a comprehensive attack-countermeasure taxonomy and a corresponding taxonomy-guided decision-support framework, complemented by existing cybersecurity tools and methodologies, to enable end-to-end cybersecurity management, from threat identification to mitigation, in manufacturing systems. Specifically, this work presents the first comprehensive manufacturing system-specific attack taxonomy to enable holistic understanding and characterization of cybersecurity threats and countermeasures. The proposed taxonomy introduces a classification scheme for threat actors and their intent and potential alterations in system behavior during threat events. It also categorizes attack methods/vectors and targets/locations, incorporates operational and system-level attack impacts, and presents a structured classification of countermeasures. The proposed taxonomy consists of six interrelated layers/dimensions: (1) *threat actors*, (2) *attack methods*, (3) *attack targets*, (4) *attack impacts*, (5) *observable anomalies*, and (6) *countermeasures*, together enabling a comprehensive analysis of cyber-physical threats in manufacturing. Unlike previous taxonomies, this framework captures the full attack chain—from adversarial motivation to observable system deviations and potential mitigation strategies. Section 2 presents the foundational requirements for taxonomy development, explains the methodology adopted, and introduces the proposed taxonomy layers. The proposed attack taxonomy for describing and documenting the complete attack chain and corresponding countermeasures is presented in Section 3. Section 4 illustrates its application by demonstrating how system anomalies can inform threat understanding and countermeasure design. Section 5 introduces a decision-support framework providing practical, hands-on guidance on how the proposed taxonomy can be integrated with other security tools, frameworks, and methodologies for threat identification, risk modeling and assessment, and selecting, designing, and deploying appropriate countermeasures. The paper concludes with a summary of key contributions in Section 6.

## 2 Taxonomy Development Methodology

This section outlines our methodology for developing a comprehensive taxonomy for cybersecurity threat attributes and countermeasures in SMS. Section 2.1 presents the foundational requirements that any standard taxonomy must satisfy. Section 2.2 discusses our approach to meet these requirements by expanding upon current taxonomies. Compared to the partial coverage of cybersecurity threat attributes and countermeasures by current taxonomies, see Table 1, this work provides novel and more comprehensive taxonomical classifications for six key threat attributes and countermeasures. The six layers of the proposed taxonomy are briefly explained in Section 2.3.

### 2.1 Key requirements

Taxonomy is the systematic classification of information to create shareable knowledge, offering a structured framework for storing and comprehending information within a specific field of study (Esmaeilian et al., 2016; Tanner & Loh, 1992). An effective manufacturing attack taxonomy should satisfy several key requirements: 1) accepted, 2) mutually exclusive, 3) comprehensible, 4) exhaustive, 5) unambiguous, 6) repeatable, 7) well-defined terminology, and 8) useful (Howard & Longstaff, 1998; Simmons et al., 2014). First, the taxonomy should be built upon well-accepted previous works. Second, different categories should be mutually exclusive with no overlap, which ensures that each entity can only be classified into one category. Third, the information provided will be clear and concise to create a classification scheme that is understood by both experts and those who are non-familiar. Fourth, the available categories within each classification will be exhaustive, facilitating a complete characterization of each attack attribute. Fifth, the taxonomy must have clearly defined classes for each of its layers. Sixth, the classification of attack incidents should be repeatable during the implementation phase. Seventh, categories should be properly defined, and those terms should comprise of accepted terminology that is compliant within the security community and industry.

Finally, taxonomy should provide insights into the cybersecurity threat characterization and help characterize the threat events. The proposed taxonomy was designed and developed to satisfy as many of these desired requirements as possible.

## 2.2 Approach

Our taxonomy development methodology incorporates a comprehensive approach depicted in Fig. 2 to align with the needs discussed in Section 2.1. First, existing manufacturing attack taxonomies were reviewed and analyzed to consolidate the scattered knowledge for attack description. IT domain cybersecurity literature, security frameworks, and industry reports were also studied to ensure the appropriate naming and definition of different attack attributes. Second, the proposed classifications of different attack attributes and their respective individual elements were compiled into a unified structure. Various research efforts emphasized developing a common language to describe and distinguish cross-domain cyber-physical attacks on SMS through attack methods, targets, and impacts. However, the attack chain in SMS begins with adversaries and their motivations, goals, and attack methods, continuing to potential attack consequences and mitigation strategies, as shown in Fig. 1. The literature overlooked a systematic and structured classification scheme for threat actors, changes in system behavior and product attributes stemming from attacks on SMS (e.g., attacks aiming to inflict physical damage), and potential countermeasures. To address the need for a more comprehensive attack taxonomy for SMS, the characterization and classification of threat actors have been incorporated into the proposed taxonomy. Additionally, new taxonomical classifications for observable anomalies during attacks and countermeasures were introduced by leveraging domain knowledge of manufacturing processes, operations, and control. The adopted methodology facilitates developing a repeatable, unambiguous, and comprehensible classification scheme by adding new, well-defined categories to the attack taxonomy knowledge base and building upon established foundations. Additionally, incorporating the missing threat attributes, i.e., threat actors and observable anomalies, into the taxonomy and presenting the first detailed classification of countermeasures makes the proposed taxonomy useful for systematically understanding and mitigating cybersecurity threats in SMS.

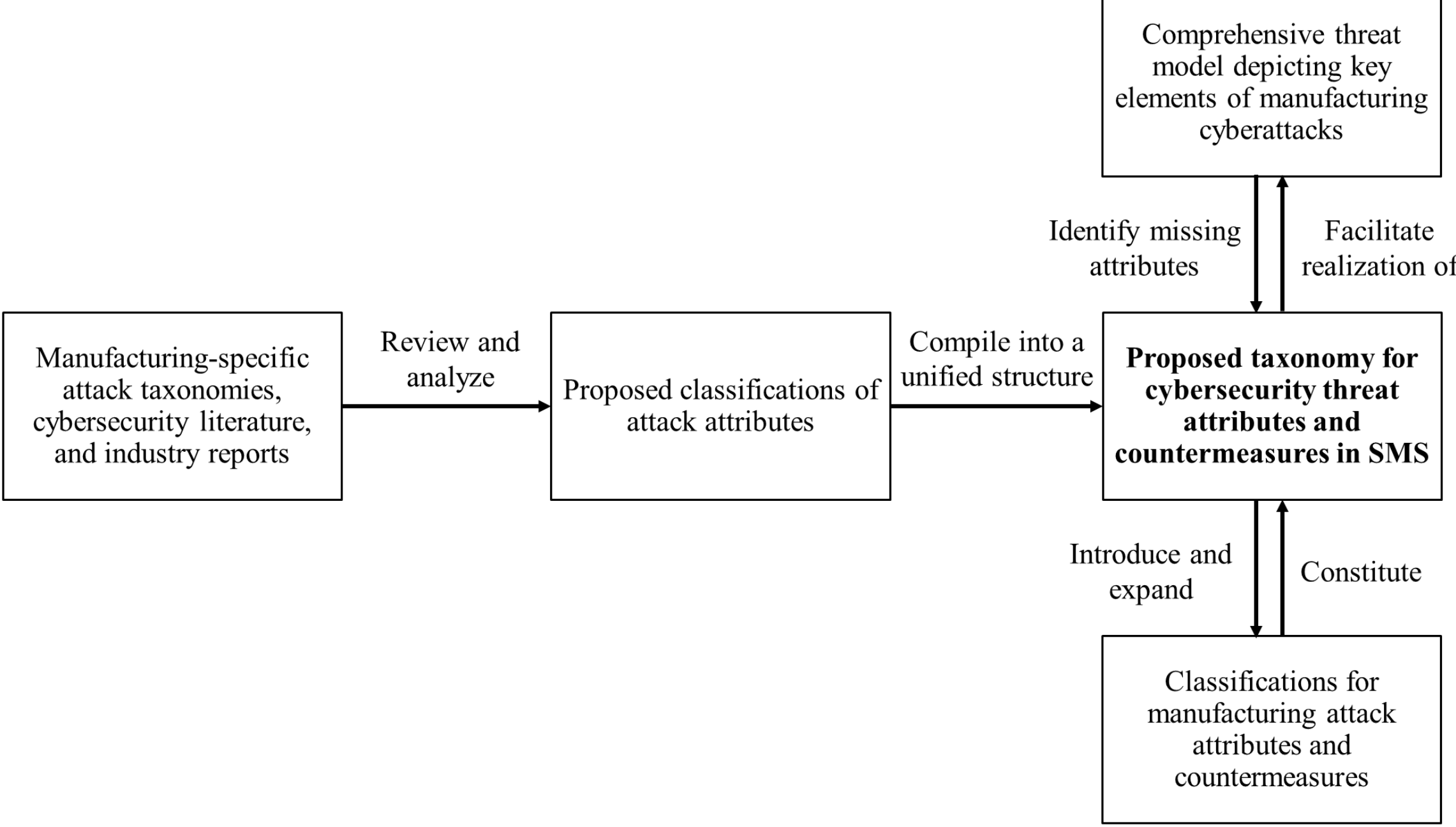

**Fig. 2**: The taxonomy development approach

### 2.3 Proposed taxonomy layers

The proposed taxonomy explains *threat actors* in its first layer to gain insight into attack methods (see Section 3.1). Threat actors may have different access levels to the system data, allowing them to execute various attacks on the system. These *attack methods* are presented in the second layer (see Section 3.2). Next, adopting the common practice of understanding a system from the threat actor's viewpoint (Elhabashy et al., 2019), the attack targets, i.e., possible system entities targeted by different attacks, are categorized in the third layer (see Section 3.3). Threat actors can harm products, processes, and the manufacturing ecosystem, depending on the target and the path of an attack. To realize this association, *attack impacts* on SMS, including the system-level and operational consequences, are characterized in the fourth layer of the taxonomy (see Section 3.4). In addition to changes in the cyber domain, cyberattacks can lead to physical changes, manifesting in product quality variations and processes. Such physical manifestations of attacks offer manufacturers an avenue to develop physical protective and detective defense measures. Hence, the *observable anomalies* are categorized in the fifth layer of the proposed taxonomy to aid manufacturers and practitioners in developing such security features (see Section 3.5). Finally, using the insights from system anomalies, the sixth layer characterizes and categorizes *countermeasures* that can be used to secure SMS (see Section 3.6). Examples of countermeasures are also provided for different phases of the SMS, which should be incorporated throughout the manufacturing workflow to mitigate cybersecurity risks.

## 3 Taxonomy for Cybersecurity Threat Attributes and Countermeasures

This section presents a comprehensive taxonomy to characterize and categorize cybersecurity threat attributes and countermeasures in SMS. It should be noted that the six layers of the proposed taxonomy are interconnected. For example, a threat actor can leverage multiple attack methods to attack different components in the manufacturing workflow. Similarly, a threat event can have several consequences, resulting in numerous observable deviations. Likewise, various countermeasures can be used to prevent or detect attacks in different phases of the digital manufacturing value chain. The six layers and their constituent elements of the proposed taxonomy are described in the following subsections.

### 3.1 Threat actors

The first layer of the taxonomy starts with identifying threat actors, which is a crucial step in understanding the security risk, capabilities of adversaries, and potential attack motives and strategies (Blank & Gallagher, 2012; Sailio et al., 2020). The threat actor can be defined as a person, group, or entity with the malicious intent of compromising an organization's security and is responsible for a cybersecurity incident, encompassing both insider and external threats. The malicious intent can range from copying confidential data onto a USB storage device to physically destroying equipment in a manufacturing facility. In reality, various threat actors display distinct attack attributes, such as attack targets, operational methods, and the infrastructure utilized to execute attacks. Hence, characterizing threat actors and their motivations can help organizations understand the threats they may face, outmaneuver adversarial tactics, and design proper countermeasures to defend against them. The following subsections briefly explain the types of threat actors and their motives.

#### *3.1.1 Types of threat actors*

Depending on the attack target and objectives, threat actors can be grouped into seven categories (Blank & Gallagher, 2012; *Canadian Centre for Cyber Security*, 2021; Sailio et al., 2020; Stouffer et al., 2013; Tuptuk & Hailes, 2018), as depicted in Fig. 3. It is worth mentioning that an individual or a group may align with multiple threat actor types. The seven types of threat actors are briefly described below.

- **Cybercriminal**. Individual or organized networks of criminals who steal sensitive product and process data or use ransomware to extort manufacturers for financial gain. They are notorious for exploiting system vulnerabilities with novel attack methods (e.g., crypto trojans) and automated tools for criminal gains. For example, a group named *XENOTIME* targeted manufacturing, oil, and gas companies in the US, Australia, and the Middle East (*Dragos*, 2020). They used the *TRITON/TRISIS* malware against an industrial safety system in 2017, crippling an industrial plant in Saudi Arabia (*Dragos*, 2020). Colonial Pipeline, the largest fuel supplier to the southeastern US, was allegedly attacked by another cybercriminal group named *DarkSide* in May 2021 and paid more than $4.4 million to resume operations after six days of disrupted operations (*Bloomberg*, 2021).

- **Nation-state actor**. These threat actors are funded and often directed by nations aiming to gather intelligence, support national interests (e.g., technology transfer, industrial espionage), and steal Intellectual Property (IP) to enhance the competencies of domestic manufacturing companies, especially those tied to military technologies. The nation-state and state-sponsored adversaries commit acts of cybercrime in favor of a government to affect a targeted nation. For example, the suspected Iranian nation-state threat actor *ITG17* (also known as *MuddyWater*) and the North Korean state-sponsored group *Lazarus* are some of the active nation-state actors targeting critical manufacturing infrastructures (*MITRE*, 2022).

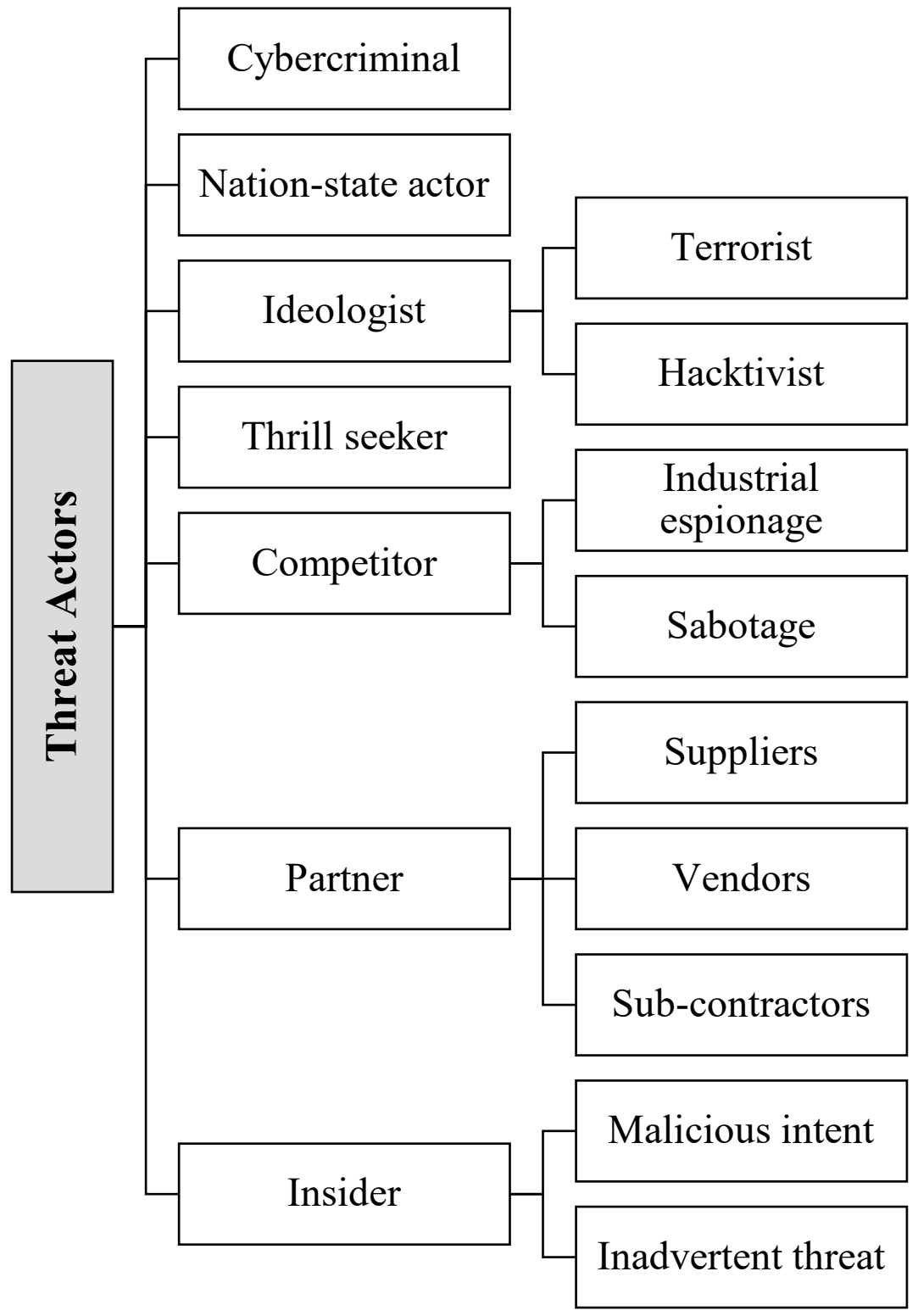

**Fig. 3**: Types of threat actors for smart manufacturing systems

- **Ideologist**. These threat actors are ideologically motivated and can be further divided into cyber-terrorist and hacktivist groups. Cyber terrorists usually aim to disrupt critical manufacturing sectors (including government facilities and public utilities) and cause harm to advance their cause. Terrorist groups can attack production systems to damage the economy or create fear in a state. On the other hand, hacktivists are driven by political ideals and focus on bringing awareness or fighting for a cause. Consequently, they often engage with their targets publicly instead of launching stealthy attacks. For example, a hacktivist group named *Anonymous* attacked Nestle, which had operations in Russia, as a form of protest against the Russian aggression in Ukraine and publicly announced the incident on Twitter (now called X) (*Fortune*, 2022).

- **Thrill seeker**. Thrill seekers (also known as script kiddies or lone wolf hackers) can launch attacks to learn, experiment, get recognition, or merely for fun. They may use tools developed by other threat actors to figure out how the system works or bugs in the computer and network systems. They usually target organizations with limited to no security, significantly disrupting the system.

- **Competitor**. Rival organizations can attack other companies for industrial espionage or damage the targeted company's reputation (also known as corporate sabotage).
- **Partner**. Partners refer to suppliers, sub-contractors, and vendors in the manufacturing supply chain with de facto insider access to various resources in manufacturing systems, such as product design and product life cycle information. Integrating an external supplier's network and data sharing with third-party distributors poses security threats for manufacturers. Partners can intentionally leak sensitive data, alter the integrity of supplied raw materials, and implant hardware backdoors into products and equipment.
- **Insider**. Insiders refer to authorized users with legitimate access to a company's assets who deliberately or accidentally abuse their access or privilege. Insiders with evil intents pose more significant threats than external threat actors because of their privileged access to organizational information, information systems, and networks, coupled with advanced organizational knowledge and the trust provided to them. Insider threat also includes the unintentional threat from undertaken actions (e.g., responding to a phishing email) or inactions (e.g., not using a strong password) without malicious intent. According to the Verizon data breach investigation report, external adversaries were responsible for 82% of the security breaches in the manufacturing sector in 2020, insiders were involved in 19% of cases, and 1% had multiple threat actors (Verizon, 2021).

#### *3.1.2 Common motives of threat actors*

Threat actors utilize their prior knowledge and access to system data and product information to launch attacks on SMS. Defending against a known threat actor is more manageable than against an unknown one. However, knowing the types of threat actors is insufficient; their motivation should be considered in the threat model for developing a robust security scheme. Some common motives for threat actors are the following (*Canadian Centre for Cyber Security*, 2021; Sailio et al., 2020):

- **Technology acquisition**. Threat actors often launch attacks to acquire cutting-edge technologies and trade secrets. In doing so, they could be motivated by political and military agendas and seek to improve their technological capabilities, economic competitiveness, and strategic advantages in various domains, including defense and intelligence.
- **Financial gain**. Adversaries can directly attack a manufacturer's financial system, encrypt system data for ransom, and/or steal proprietary data and sell it for profit.
- **Notoriety**. Some threat actors publicize attacks to seek reputation, status, or mere attention.
- **Revenge**. Revenge is another common attack motivation, especially for disgruntled employees and nation-state actors.
- **Overlap of motivation**. A threat actor may have multiple incentives to attack SMS. For example, a disgruntled employee with a revenge mindset may share intimate knowledge about an organization or confidential data with other threat actors for financial gain.

It is beneficial for the cybersecurity risk assessment process and forensic analysis following threat events to understand and characterize threat actors as well as their underlying motives and aims. With that, manufacturers can reduce the range of potential threat actors for their specific environment based on organizational category and product line. Additionally, they can design and develop appropriate countermeasures and allocate resources to prevent or mitigate the acts of respective threat actors. For example, suppose a manufacturer is aware that a nation-state actor is targeting it. In that case, it can take steps to protect itself from more sophisticated and targeted attacks, such as strengthening access controls and additional monitoring for unusual activity. In contrast, if a company is aware that a gang of “script kiddies” (unskilled individuals who rely on pre-existing scripts or programs developed by others (*Cybersecurity Magazine*, 2023)) is targeting it, it may be able to defend against such threats by implementing more basic control measures, requiring less investment.

## 3.2 Attack methods

The second layer of the taxonomy categorizes attack methods (also known as attack vectors), presenting the specific tools and techniques that threat actors use to access the manufacturing ecosystem and execute the attack. As shown in Fig. 4, threat actors can employ different attack methods based on their prior knowledge, access to system data, and available resources to fulfill their objectives, including stealing confidential information, reducing outgoing product quality, altering a product's design intent, or disrupting manufacturing processes. Additionally, various attack methods can be associated with specific attack targets. Hence, understanding these methods can help develop adequate defenses and facilitate incident response and recovery. Depending on the domain where an attack is initiated, attack methods can be broadly categorized into three groups: 1) cyber domain, 2) cyber-physical domain, and 3) physical domain. Cyber-domain attack methods are primarily launched over the network communication system, physical domain attack methods are initiated after gaining physical access to the manufacturing asset, and cyber-physical attack methods can leverage both cyber and physical means for attack execution. Different attack methods within the three groups are shown in Fig. 4 and briefly described below with examples of potential attack targets, i.e., origins and sub-origins (discussed in Section 3.3).

### *3.2.1 Cyber domain attack methods*

- **Denial of Service**. Denial of service (DoS) is typically accomplished by flooding the target with several requests to overload systems and prevent the execution of legitimate requests. Conversely, a Distributed Denial of Service (DDoS) attack occurs when multiple machines work together to attack one target. Threat actors can target the communication network (origin) for both attacks.
- **Man-in-the-middle attack**. This attack aims to interrupt the data flow between different communication endpoints (Bhushan et al., 2017). Threat actors can launch this attack at the communication network (origin) to obtain confidential instructions, commands, and information (sub-origins) pertinent to manufacturing operations.
- **Compromised software/logic**. Threat actors can also compromise the software/logic (origin) used in SMS. For example, they can target commonly used inspection, design, and management software (sub-origins).
- **False data injection**. False data injection attack refers to inserting malicious input into a system (such as a web application) to conceal undetected errors, forcing the system to execute specific malicious commands (Ahmed & Pathan, 2020). To implement this attack, threat actors can target the communication network (origin) to provide malicious instructions, commands, or information (sub-origins).
- **Privilege escalation**. Privilege escalation refers to a scenario where an insider (threat actor) can gain unauthorized and unregulated permissions to a system or network (*MITRE*, 2022). For example, threat actors can use this technique to enter the communication network system (origin) and access confidential digital files in cloud storage (sub-origin).
- **Replay attack**. A network attack in which valid data transmission is fraudulently repeated or delayed (El Abbadi & Jamouli, 2021). This attack also targets the communication network (origin), where adversaries may manipulate digital files (sub-origin) containing *in situ* monitoring data from the production process.
- **Web attack**. Web attack is a broad class of methods referring to phishing, ransomware, malware, password theft, and spoofing attacks targeted at the communication network (origin).
- **Zero-day attack**. This attack exploits vulnerabilities that have not been publicly revealed (Bilge & Dumitras, 2012). Adversaries can exploit novel software/hardware vulnerabilities before developers create a patch to fix the zero-day (FireEye, 2023).

- **Tampering with digital file/G-code/toolpath**. This attack refers to an intentional but unauthorized modification of CAD files, G-codes, and toolpath files (NIST, 2025). Threat actors can alter digital files to manipulate process and post-processing parameters (sub-origins) during production.

### *3.2.2 Cyber-physical domain attack methods*

- **Social Engineering**. Social engineering is a social/psychological process that threat actors can use to manipulate human workers (sub-origin) to gain confidential information about a targeted organization (Thornburgh, 2004).

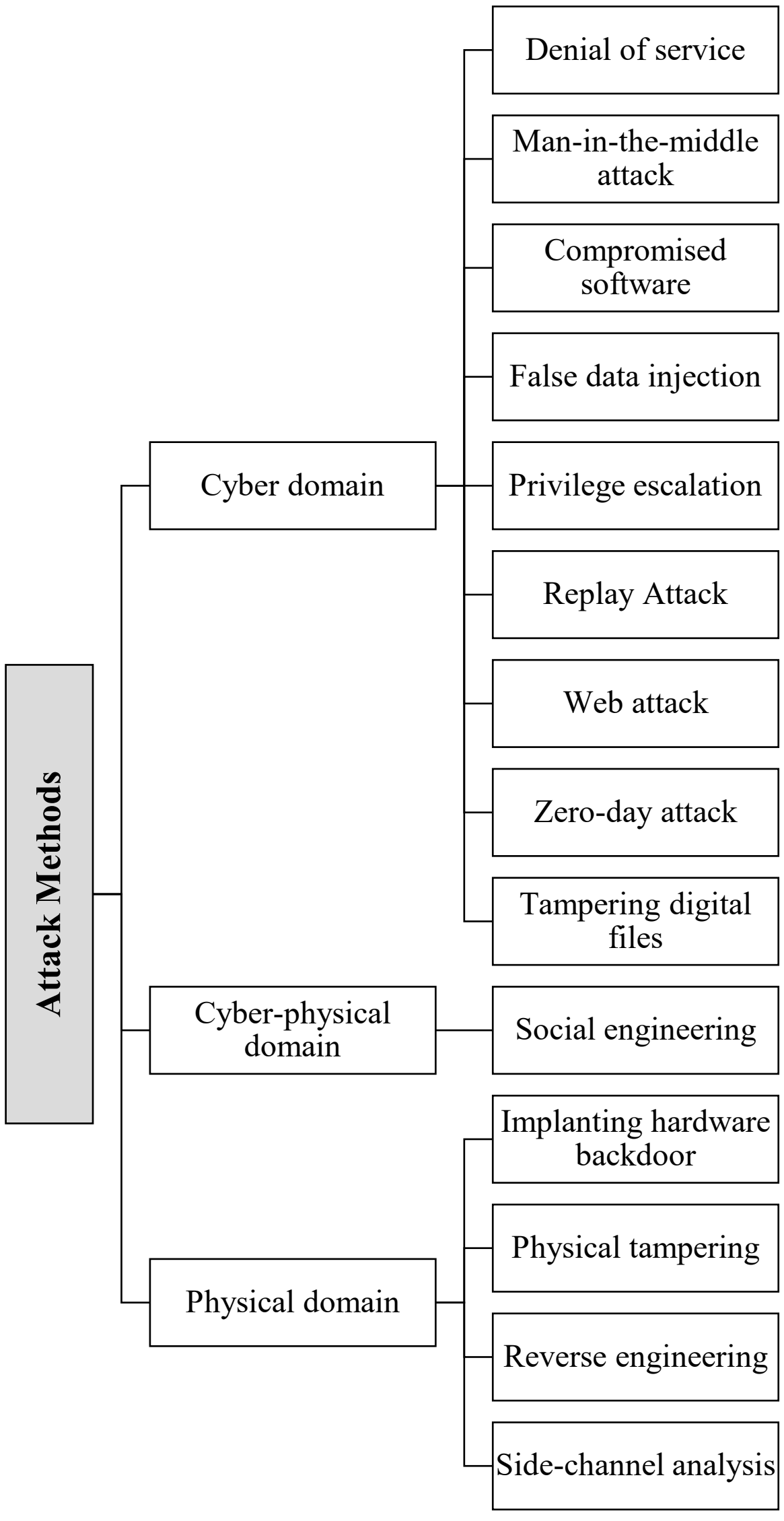

**Fig. 4**: Classification of attack methods/vectors

### *3.2.3 Physical domain attack methods*

- **Implanting hardware backdoor**. Manufacturing machines can have embedded unauthorized commands/code in the firmware that can allow threat actors (usually state-based actors in this case) to control the device after installation (ICS-CERT, 2016; *The Guardian*, 2012). For example, the firmware of a 3D printer interprets the tool path files and gives an actuation command to different motors for nozzle movement and filament extrusion. Tampered firmware can misinterpret tool path files and/or send inaccurate actuation commands to motors, eventually printing defective parts.

- **Physical tampering**. Tampering with a physical part means the unauthorized malicious modification of a part or system. This attack method aims to reach a physical component of the system and can be categorized into two forms: tampering with a) hardware and b) raw material. For physical tampering of the hardware (origin), potential attack targets can be the controller, actuator, and sensor (sub-origins).
- **Reverse engineering**. This attack refers to producing counterfeit parts after analyzing the functionality and behavior of the original part (Mahesh et al., 2021). Adversaries can reverse engineer specific manufactured components (origin), alter their geometry and structure (sub-origin), and inject counterfeit parts with compromised functionality into the product supply chain.
- **Side-channel analysis.** This type of attack focuses on reconstructing IPs, such as the 3D model of a part, by observing the acoustic, vibration, and magnetic emissions (sub-origins) during discrete part production (Song et al., 2016). For example, threat actors can use mobile devices such as smartphones to measure these emissions from the manufacturing environment (origin) in close proximity to machines.

## 3.3 Attack targets

The third layer of the proposed taxonomy consists of attack targets in the digital manufacturing value chain. Attack target refers to the specific system, device, or network to which an attacker seeks to exploit or gain unauthorized access. Characterizing the attack target allows manufacturers to identify and prioritize potential entry points for threat actors and vulnerabilities. Understanding the attack target can also offer insights into determining whether an attack is a standalone incident or part of a more extensive campaign. This paper classifies the attack target into two sub-categories: *origins* and *sub-origins*. The *origin* refers to a macro-category, representing where adversaries can initiate the attack, and the *sub-origin* provides a more granular understanding of the attack targets. To illustrate, threat actors can target the software (origin) and, more specifically, aim for the design software (e.g., SolidWorks). The primary attack targets, i.e., origins, are briefly described below, and the corresponding sub-origins are presented in Fig. 5.

- **Hardware**. It refers to physical components in SMS. This work considers the controller, actuator, and sensor (sub-origins) as hardware components, all of which can be affected by physical tampering (attack method).
- **Software**. In this study, the macro attack target is articulated in four sub-origins, and all of them (see Fig. 5) can be affected by the compromised software/logic attack (attack method) by an external adversary (threat actor). For example, a threat actor can exploit vulnerabilities in SolidWorks, a design software, to tamper with the GD&T data in a product design.
- **Cloud storage, cloud services, and communication networks**. Cloud storage and the network system are prime targets of adversaries for attacking digital files and instructions (sub-origins) pertinent to manufacturing operations, such as CAD files, computer-aided process planning files, tool paths, and GD&T information. An external adversary can access these files and instructions through privilege escalation or a web attack (attack methods) (H.-T. Lee et al., 2016; M. Wu & Moon, 2017).
- **Manufacturing environment**. Includes all the elements in the SMS that are not directly involved in the production process. The manufacturing environment includes but is not limited to material properties, energy consumption, and emissions (sub-origins). Adversaries can physically tamper (attack method) with the manufacturing environment instead of the manufacturing process to disrupt production. For example, the metal powder properties can be tampered with instead of altering the 3D printer in a powder bed fusion AM process.
- **Component/product**. Adversaries can attack the geometry and structure, process parameters, and post-process parameters (sub-origins) to manipulate the design intent of a component/product (origin). For example, adversaries can attack machining process parameters such as cutting speed, feed rate, and depth of cut to alter the geometric features of a product. In doing so, they can tamper with digital files/G-codes/toolpaths (attack methods).

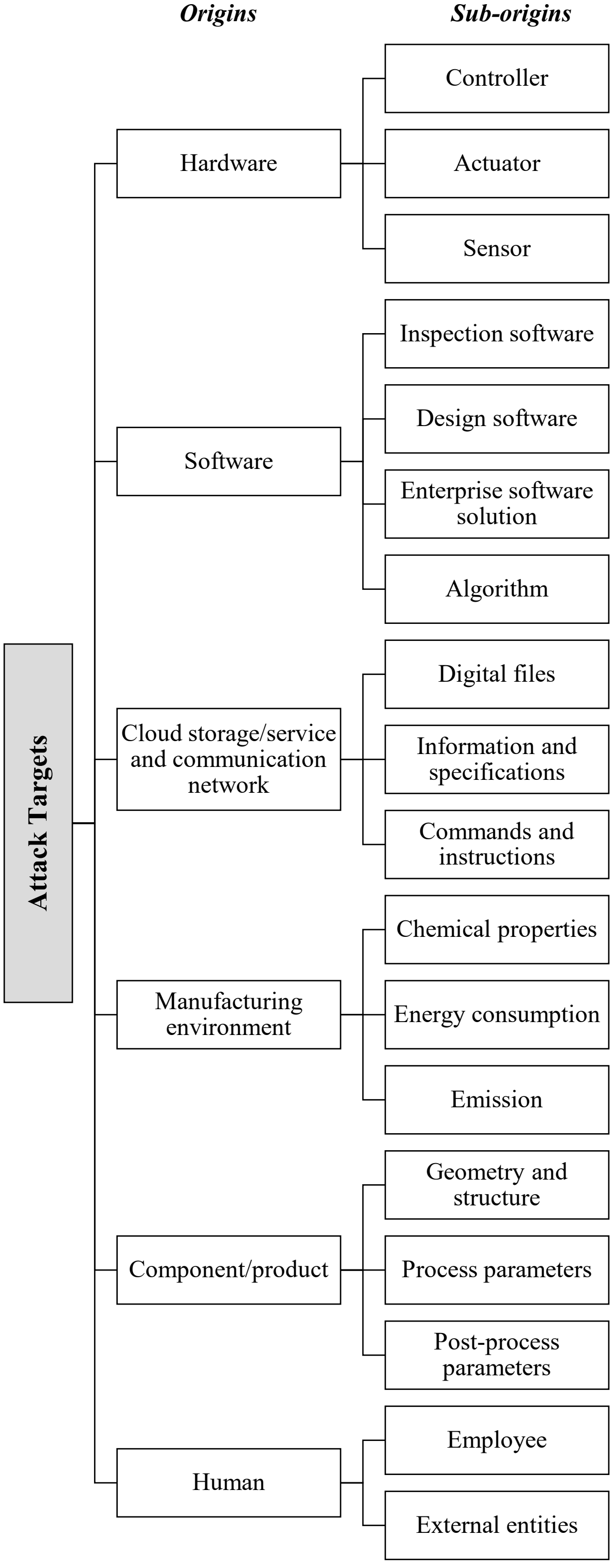

**Fig. 5**: Attack targets in the digital manufacturing value chain

## 3.4 Attack impacts

The fourth layer of the proposed taxonomy categorizes potential impacts on the manufacturing ecosystem arising from successful cyberattacks. Up to the attack target layer in the taxonomy, the attack chain was explored from a threat actor's viewpoint. This approach can help assess potential attack strategies and evaluate how threat actors can launch attacks on one or multiple digital manufacturing value chain elements. However, the information on the threat actors' intent and attack methods can be limited, and only a subset of the attacks can pose a significant risk to the SMS. In essence, manufacturers, especially small and medium-sized enterprises, and practitioners may be more interested in analyzing attack impacts on SMS for the impact-oriented security risk assessment. Hence, moving forward, the attack chain is investigated from the manufacturers' viewpoint to assess the potential consequences of attacks on manufacturing systems.

Attack impacts/consequences can be primarily grouped into four categories: 1) theft of confidential information and monetary loss, 2) impact on the system, 3) operational impacts, and 4) product-related impacts. The four categories of attack impacts are described below.

### *3.4.1 Theft of confidential information and monetary loss*

In manufacturing systems, cyberattacks threaten data confidentiality, integrity, and availability (also known as the CIA triad in IT security). Confidentiality emphasizes disclosing data only to the authorized person, integrity focuses on protecting that data from malicious modifications, and availability ensures that the data is always accessible to authorized entities (Tuptuk & Hailes, 2018). The confidentiality breach can lead to strategic and financial losses in manufacturing. Threat actors can use ransomware to encrypt important files and data of the victim organization and demand a ransom to restore the victim's access. They can also steal confidential manufacturing processes and product data for financial gains. For example, ransomware attacks and data theft accounted for 37% of all attacks on manufacturing in 2020 (IBM, 2021). Apart from stealing data from the cyber domain, IP can be stolen by analyzing manufacturing process outcomes known as side-channels (see Section 3.2).

### *3.4.2 System-level impacts*

In contrast to the IT domain, integrity and data availability are also associated with the system reliability, safety, and maintainability of physical manufacturing systems. Consequently, product and process-oriented cyberattacks can significantly harm or sabotage the manufacturing ecosystem. To illustrate, modern Fused Deposition Modelling (FDM) 3D printers are equipped with temperature sensors to gauge the nozzle temperature. The sensor data maintains a specific temperature for melting the filament for extrusion. The printer may fail to maintain the required temperature for sustained material extrusion if the sensor data becomes unavailable. The effect could be even more disastrous if the integrity of that data is tampered with. For example, the communication protocol (origin) of a FlashForge printer was attacked in 2020 to tamper with the machine firmware, i.e., a digital file (sub-origin), to eliminate the printer's maximum temperature constraint (viz., violation of data integrity) (McInerney, 2020). This attack increased the temperature of the heating element to the point of catching fire, ending up weaponizing a benign 3D printer. The "massive" physical damage of a German steel mill (R. M. Lee et al., 2014) is another example of system-level impact.

### *3.4.3 Operational impacts*

Cyberattacks, especially process-oriented attacks, can have significant operational impacts. In distributed and collaborative manufacturing systems, production control operations include demand forecasting, capacity planning, and production planning operations such as Material Requirements Planning (MRP), scheduling, and inventory control. With the broad adoption of automation technologies and product life-cycle management tools, most operations are automated, digitally monitored, and controlled through Enterprise Resource Planning (ERP) software. This automation comes with its own vulnerabilities. For example, several significant weaknesses were found in Oracle WebLogic, one of the most widely used ERP solutions, allowing threat actors to take control of the server and tamper with operational data (*CVE-2020-14750*, 2020). Compromising such ERP applications (sub-origin) through a denial of service (attack method) alone can cause significant production downtime (attack impact). Additionally, various

Automated Guided Vehicles (AGVs) are used for transportation and material handling in manufacturing facilities and warehouses, which are pre-programmed to avoid collisions and operational deadlock (no movement). Threat actors can manipulate such control algorithms (sub-origin) to disrupt their operation schedule (attack impact), creating backlogs in material transfer. Production downtime due to an attack can also be considered an operational impact. Operational impacts can have cascading consequences on the overall manufacturing supply chain. For example, Toyota had to shut down fourteen factories in Japan in 2022 when one of its suppliers was hit by cyberattacks and failed to deliver parts, halting one-third of Toyota's global production for one day (*CNN Business*, 2022).

#### *3.4.4 Product-related impacts*

On a product level, cyberattacks can degrade material properties (e.g., thermal conductivity), product geometry (e.g., size and shape), dimensional tolerances (e.g., surface roughness), and the interface with other components (e.g., assembly operation). For example, threat actors can tamper with the tool path file (sub-origin) of a CNC machined part during the file transfer (origin) to the controller computer, which can significantly reduce the printed parts' tensile strength and/or functional performance (attack impact) (Wells et al., 2014).

It is worth mentioning that attack impacts are not mutually exclusive, and coordinated attacks can simultaneously cause multiple damages. To illustrate, an adversary may breach the cloud storage, access product design files, and obtain information about the manufacturer's inspection strategy to know which parts are monitored. Assume the manufacturer has a post-production inspection plan to monitor 10% of the parts, i.e., inspect every tenth part. Information regarding this sampling strategy can later be used to launch product-oriented attacks (such as changing the geometric features of a part) on the parts that will not be inspected (Rahman & Shafae, 2022; Shafae et al., 2019). In this example, the attack can result in the theft of confidential information, such as the proprietary product design file, and product-related impacts, such as reduced mechanical strength and functional performance of the final product from malicious alteration of geometric features in the design file.

### 3.5 Observable anomalies

The fifth layer of the proposed taxonomy categorizes observable anomalies in the manufacturing system associated with the above-mentioned attack impacts. Cyberattacks can affect manufacturing systems in various ways depending on the adversarial goal, attack method, and target. These attacks will result in specific anomalies in the system's behavior, allowing the manufacturing community to detect those attacks. Anomaly-based (also known as behavior-based) attack detection approaches, which look for system anomalies for the "expected" or "normal" behavior of the system, are commonly used in IT systems (Khraisat et al., 2019; Tuptuk & Hailes, 2018). This approach utilizes various techniques, including statistical methods and machine learning, to define or learn the "normal" system behavior with no attacks and can distinguish anomalies arising from attacks (Khraisat et al., 2019). While this approach is being implemented in manufacturing systems for attack detection in information and communication systems, little attention has been given to leveraging the information provided by the actual physical behavior of production systems. Product and process-oriented cyberattacks also manifest signs of abnormal behavior in process dynamics. For example, adversaries can send malware as email attachments to the employees of the target organization. These anomalous communication packets can contain ransomware to encrypt specific files in the system once opened and executed by an employee. This attack will only manifest anomalies in data in the cyber domain. However, the malware could be designed to change the CAD or G-code files to maliciously alter the geometric features of a machined part, which will also inflict changes in the machining process and the product geometry. Anomalies from the machining process's nominal behavior and the product's nominal geometry can be leveraged to design robust anomaly-based approaches to prevent and detect attacks on manufacturing systems. The following sub-sections briefly discuss observable anomalies in manufacturing systems, highlight the difference between attack-induced anomalies and process variation, and classify anomalies.

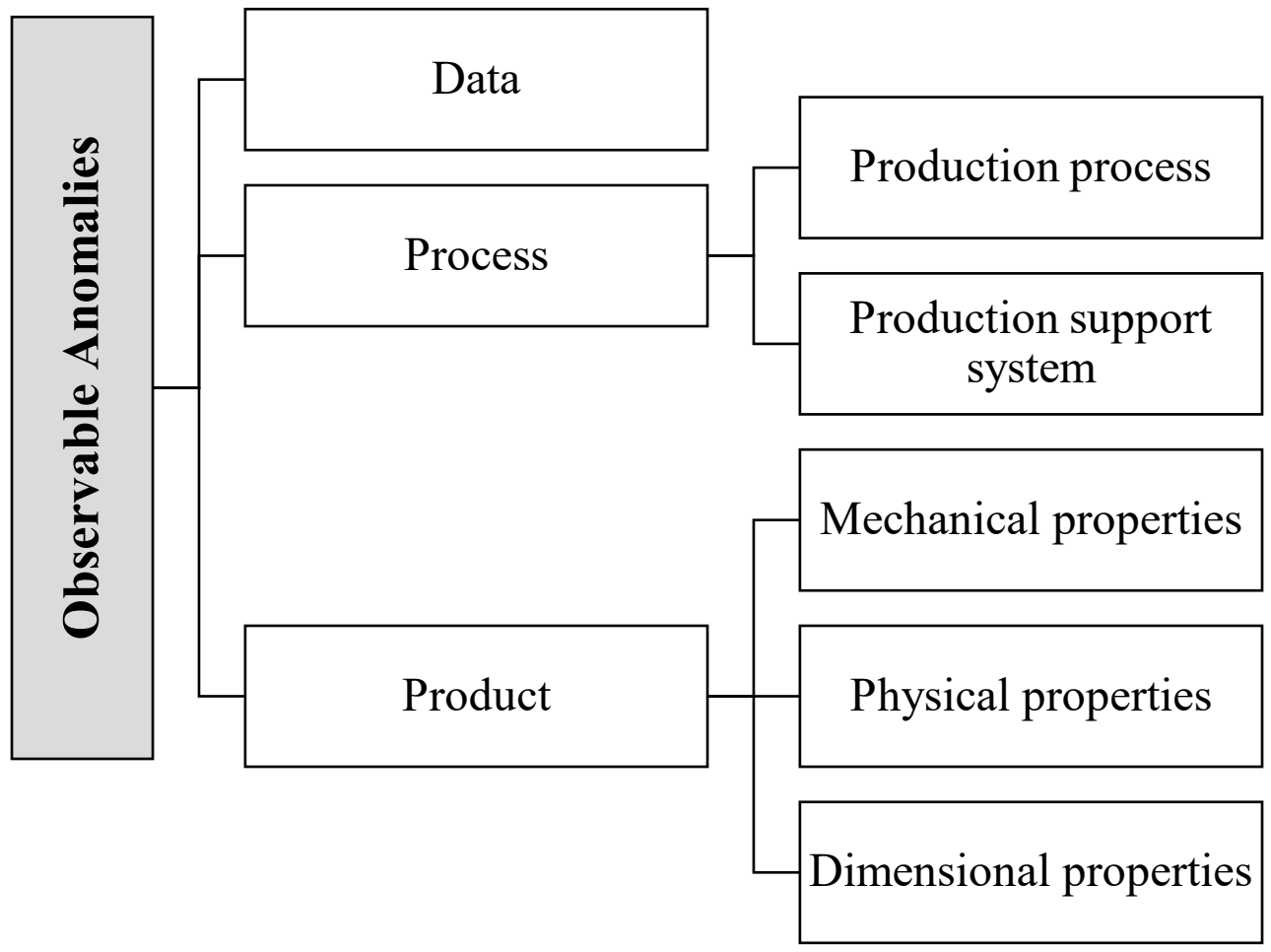

**Fig. 6**: Potential anomalies observed during cyberattacks in manufacturing systems

#### *3.5.1 Levels of observable anomalies*

Considering different attacks and their impacts, potential anomalies in manufacturing systems can be observed on three levels: 1) data, 2) process, and 3) product. Data anomalies refer to changes in commands, instructions, and digital files in the host computers and network links. As shown in Fig. 6, process anomalies can be categorized as alterations in the production process (such as machining sequence, tool selection, and process parameters) and production support systems (such as computer-aided process planning and material handling systems). Potential anomalies in the product can be further classified into mechanical properties (such as strength and hardness), physical properties (such as thermal and electrical properties), and dimensional properties (such as dimensions, tolerances, and geometric features). A shift from the nominal characteristics of data, process, and product is considered an anomaly. Note that these anomalies are not mutually exclusive. Attacks targeting the cyber domain (such as false data injection) may only inflict anomalies in the cyber domain. In contrast, attacks against the physical manufacturing domain (such as hardware tampering) will manifest in product and process anomalies. In comparison, cross-domain attacks can simultaneously cause data, product, and process anomalies. For instance, threat actors can target the control algorithm or software in the cyber domain to affect the material handling system in the physical domain, which will cause anomalies in both the data and the process level.

#### *3.5.2 Process and product variations vs. anomalies due to an attack*

Differentiating malicious and non-malicious incidents in an extensive, interconnected manufacturing system can be overwhelming. In reality, attributes of physical processes and manufactured parts always show some variation due to natural causes or noise, known as common cause variations. Common causes include measurement error, normal wear and tear of equipment, variability in machine settings, and changes in the production environment. In contrast, unexpected variations, also known as special cause variations, are caused by specific causes (also known as assignable causes in the quality control field), such as power surges and machine malfunction (Shewhart, 1930). Variations due to assignable causes and cyberattacks can have similar effects, which makes distinguishing them challenging. However, anomalies from assignable causes and cyberattacks need to be distinguished from variations due to noise.

While variations can be found in the process due to natural causes, such anomalies are ubiquitous across the production process (Graves et al., 2021). On the other hand, anomalies followed by attacks or assignable causes are more localized and might be observed sporadically. Due to the sequential nature of the manufacturing value chain, anomalies in process and product incurred at an early stage are likely to be amplified along the following stages. A significant deviation observed at a specific stage may also result from accumulating minor anomalies from previous stages. Hence, the correlation of anomalies across the different stages can significantly improve cyberattack detectability and reduce false alarms. Additionally, specific process variables may influence additional variables with a time delay due to the dynamic nature of manufacturing processes (Sánchez-Fernández et al., 2018). For example, the cutting temperature during machining will increase after the cutting fluid flow is interrupted; the change may not be instantaneous. That said, the characterization of anomalies requires the knowledge of temporality (time and frequency of occurrence), multiplicity (number of components or stages involved), and the domain (physical or cyber) of the deviation (Lopez et al., 2017).

#### *3.5.3 Classification of anomalies*

Anomalies can be classified into five categories based on temporality, multiplicity, and domain (Lopez et al., 2017). The categories are: 1) instantaneous, 2) evolving, 3) communication, 4) event-based, and 5) integration anomalies.

- **Instantaneous anomalies**. Instantaneous anomalies manifest in the system without prior indication and may originate in the cyber domain (such as unusual data packets) or physical domain (such as increased cutting force during machining).
- **Evolving anomalies**. Refers to anomalies that are dynamic in nature and manifest through process evolution. These anomalies can originate in single or multiple components and generally change over time.
- **Communication anomalies**. Anomalies in the communication network include receiving faulty data packets, unusual data traffic, and missing communication. Communication anomalies can be observed in single or multiple components and detected through the evolution of observations.
- **Event-based anomalies**. This category refers to anomalies manifested as the missing of certain events (such as machining sequences and actuator actions) or the unexpected occurrence of events. Event-based anomalies can only be detected through dynamic observations of the system.
- **Integration anomalies**. Integration anomalies can be observed in both cyber and physical domains, representing an accumulation of minor anomalies. These anomalies manifest in a system where the final outcome is anomalous despite individual system components seeming to function normally.

Among these five deviation categories, multiple anomaly classes can be found in data, process, and product. For example, dimensions and geometric features in a product can deviate from the nominal specification due to cyberattacks during the multi-stage production process. Adversaries can tamper with design files, G-codes, and toolpath files using product-oriented attacks or physically alter machine set-up and inspection processes. Consequently, anomalies can originate from both cyber and physical domains. For product-oriented attacks, the alteration in dimension and geometric features in a machined part can occur in one instance or via small changes across the machining process (temporality attribute of the deviation) (Shafae et al., 2019). As dimensional and geometric anomalies can be instantaneous and evolving depending on the attack strategy, the in-situ and post-production inspection systems should be designed accordingly (Rahman & Shafae, 2022).

### 3.6 Countermeasures

Potential countermeasures against the cybersecurity threat are categorized and discussed in the sixth layer of the proposed taxonomy. Following the attack chain, attack consequences and their manifestation through data, process, and product anomalies have been discussed. Knowledge of these anomalies can now guide the development of appropriate control measures for preventing, detecting, and mitigating cyberattacks. While the previous layers of the proposed taxonomy offer insights into the threat landscape through threat attribute characterization, the taxonomical classification of countermeasures can assist organizations in selecting the most suitable countermeasures for their risk profiles. The following subsections propose a classification of countermeasures and explain it through examples of potential countermeasures.

#### *3.6.1 Countermeasure classification*

As presented in Fig. 7, this work classifies countermeasures based on three key features: a) activity level, b) deployment location, and c) attack response strategy. The activity level represents whether a countermeasure is designed to 1) prevent threat events, 2) detect threat events when they occur, or 3) mitigate the threat post-occurrence. The deployment location refers to the location in the digital manufacturing value chain where the countermeasure is deployed. For example, the workflow for discrete part production starts with designing a product using Computer-Aided Design (CAD) software. The design files are later converted to machine-readable codes and transferred to the controller computers connected to the machines in the production facility. The machine firmware reads and executes the machine code. Numerous process variables are frequently measured in situ and with machine-embedded sensors before being sent to edge and/or cloud computing devices for real-time monitoring and control of production processes. Different post-production inspection techniques are also used to validate the outgoing product quality. Appropriate countermeasures can be deployed at all these stages of the value chain. The deployment location is classified into four phases for generality: 1) product design, 2) production, 3) post-production inspection, and 4) distributed over the value chain. Finally, countermeasures can be characterized by the attack response strategy built into them. More specifically, countermeasures can be designed and developed to 1) improve cognitive awareness as a preemptive action, 2) produce alerts and reports during threat events, 3) provide evidence to diagnose threat events, and 4) reconfigure the system for threat mitigation.

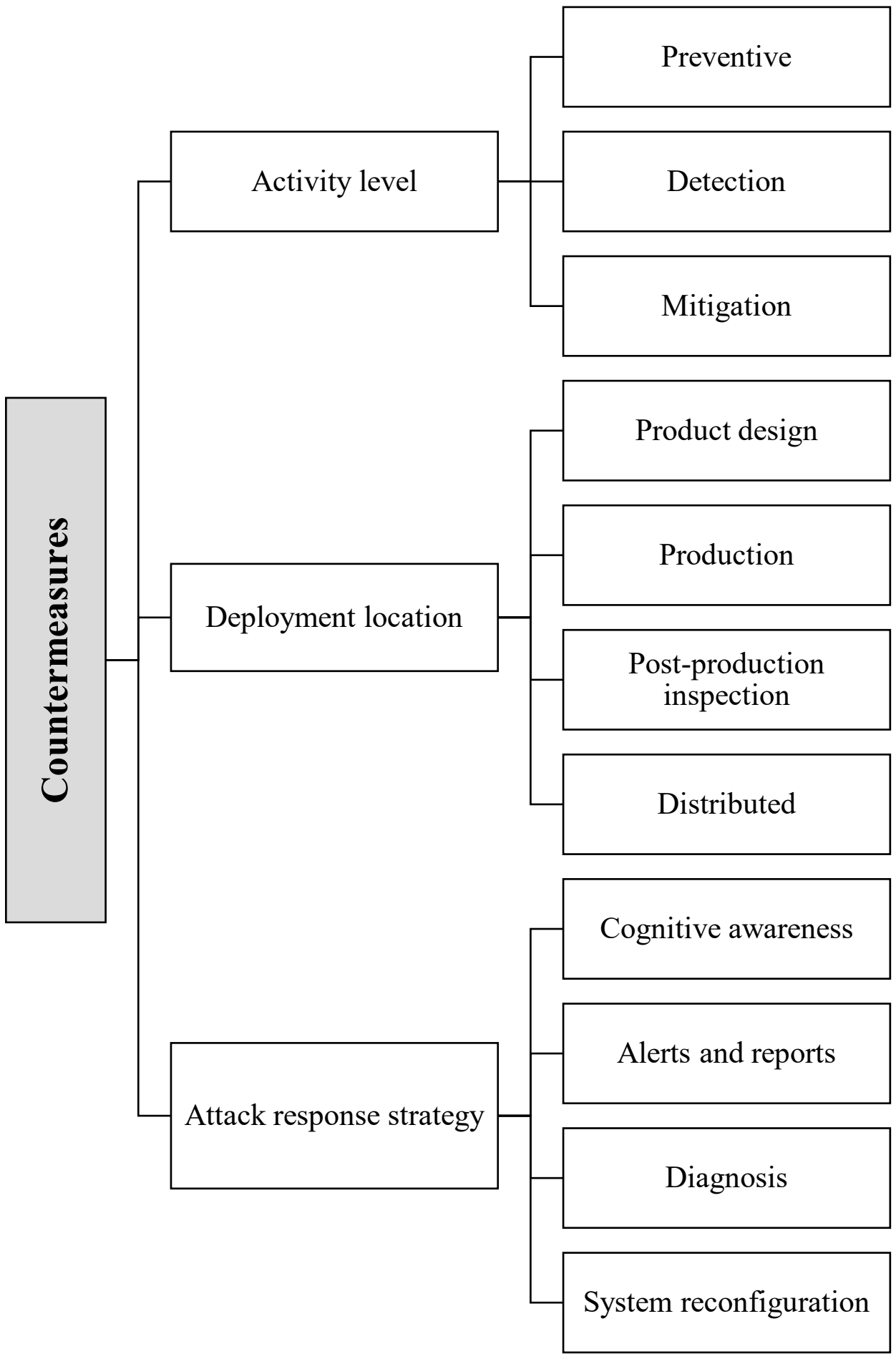

**Fig. 7**: Classification of countermeasures against cyberattacks in manufacturing systems

### *3.6.2 Examples of potential countermeasures*

This section offers some examples of potential countermeasures and explains how they fit into the proposed taxonomical classification. As described in Section 3.5, attacks on manufacturing systems can inflict changes in data, processes, and products. Common cyber-domain security solutions include host and network-based Intrusion

Detection Systems (IDS), which focus on data anomalies. However, the scope of protection must include both IT systems that process data and operational technologies that operate machinery (*The White House*, 2021), which calls for defense measures to detect anomalies in processes and products. If adversaries bypass the cyber domain security, such as the IDS, the physical changes inflicted by the attack can still be detected in the process and product. Acknowledging the need for countermeasures in both cyber and physical forms, several effective countermeasures are briefly described below, and they are characterized based on the proposed taxonomical classification in Table 2.

- **Intrusion Detection Systems (IDS)**. IDS detects unauthorized network traffic and sensitive data access, misuse, and alteration (Zhang et al., 2011). It detects suspicious network traffic, system activity, and user behavior via signature-based detection, anomaly detection, behavior-based detection, and correlation-based detection. Once an IDS detects an intrusion, it can alert the security staff to stop the attack and create reports for further investigation and analysis.
- **Network segmentation**. Network segmentation divides the factory network into subnetworks to safeguard each separately. It reduces the attack surface and limits cyberattacks. Network segmentation prevents attackers from moving laterally and accessing critical systems and data. Network segmentation enforces access rules and limits network traffic across various parts of the network, reducing the likelihood of a cyberattack.
- **Cybersecurity training**. Cybersecurity training consists of teaching individuals or organizations how to safeguard computer systems, networks, and sensitive information from unauthorized access, use, disclosure, disruption, modification, or destruction through a variety of tools, such as policies and procedures. Security awareness and skills training are among the five most effective protective measures against cyberattacks on manufacturing systems, according to the Verizon data breach investigation report (2021) (Verizon, 2021).
- **Watermarking.** The watermarking technique is a protective defense measure that embeds inseparable and hidden information in the design file that only authorized users can use to authenticate the files (Mahesh et al., 2021).
- **Embedding QR codes.** Multiple segments of QR codes can be embedded in the CAD model in additively manufactured parts for product authentication (Chen et al., 2019). QR codes can be placed in pre-defined strategic orientations only known to authorized persons. Consequently, even if adversaries get access to the digital file, they will end up manufacturing a part showing the counterfeit mark. Embedding such codes increases detectability against counterfeit parts.
- **Physically Unclonable Functions (PUFs).** The PUFs can be used as a protective technique against manufacturing process tampering (Maes & Verbauwhede, 2010). The physical characteristics of manufacturing processes and their surrounding environment can be used to create unique fingerprints, which can validate the integrity of the manufacturing process (Mehdi & Starly, 2020; Sandborn et al., 2021).
- **Physics-informed detection methods.** Insights on process physics and potential anomalies can be leveraged to develop an attack detection mechanism at the machining process level. Malicious changes in digital files and process parameters will demonstrate anomalous behavior in process dynamics. These changes can be detected by observing in-process variables (such as acoustics, vibration, and power consumption) known as side channels to the physical process. For example, spindle power can be observed during CNC turning operations to assess the integrity of the machining process and the machined part (Lin et al., 2023; Rahman & Shafae, 2022).

It should be noted that the relationship between threat attributes and countermeasures is not always linear; a single countermeasure may protect against numerous phases of the attack chain, while the specific countermeasure may not be effective against all forms of attack. Different countermeasures have varying optimal deployment strategies, require varying degrees of coordination with other defenses, and respond differently to attacks. Comprehending the scope, limitations, and constraints of available and potential defense measures and adding countermeasures to the taxonomy can ensure the seamless integration of threat realization and mitigation (Rahman et al., 2023).

**Table 2:** Example countermeasures and their characterization following the proposed classification scheme

| Countermeasure | Activity level | Deployment location | Attack response strategy |
|---|---|---|---|
| Intrusion detection systems | Detection | Distributed | Alerts |
| Network segmentation | Mitigation | Distributed | System reconfiguration |
| Cybersecurity training | Protective | Distributed | Cognitive awareness |
| Watermarking | Protective | Product design | Diagnosis |
| Embedding QR codes | Protective, Detective | Product design | Alert, diagnosis |
| Physically Unclonable Functions | Detective | Production | Alerts |
| Physics-informed detection methods | Detective | Production | Alerts, diagnosis |

# 4 Taxonomy Implementation Examples

Two examples are presented in Sections 4.1 and 4.2 to show the implementation of the proposed taxonomy. These examples are carefully selected to showcase the value of having a holistic taxonomy and help understand the connections between the proposed layers. From the threat-oriented cybersecurity risk assessment viewpoint, the first example provides the threat actor's perspective of the attack chain. It shows attack paths to reach a specific attack target for inflicting damage to the attacked system. In contrast, the second example provides the manufacturer's perspective, highlighting attack consequences, potential system anomalies due to the attacks, and suitable countermeasures to detect and/or prevent the attacks. Finally, several cyber-physical attack incidents and case studies are analyzed using the proposed taxonomy in Section 4.3.

## 4.1 Scenario 1: An insider stealing confidential information or sabotaging operations

In 2018, a Tesla employee reportedly conducted "extensive and damaging sabotage" to their operations, including direct code changes to the Tesla manufacturing operating system using fake usernames (*CNBC*, 2018). The employee also traded highly sensitive Tesla data to unidentified third parties. The stated motivation for this attack was that the employee wanted a promotion but did not receive it. Likewise, A disgruntled employee can turn against a manufacturing company out of a personal grudge or for financial gain. Insiders often have unrestricted access to physical manufacturing assets and sensitive information regarding product and process specifications and the company's business operations. With such access, an insider threat can (1) steal confidential information and (2) sabotage operations targeting production processes, products, and the manufacturing ecosystem. They can exploit the lack of authentication and poor security policies to read, modify, and/or maliciously download sensitive and confidential data about products and processes. They can also tamper with production processes on the shop floor and/or disrupt operations across the manufacturing value chain.

### *4.1.1 Theft of intellectual property*

In this illustrative attack scenario, the specific *attack target* could be the information on products and production processes, which is identified as a sub-origin in the taxonomy. The threat actor could first target the cloud storage (origin) to access the desired information, as sensitive information is often shared or saved in cloud storage. Cloud storage and services are often vulnerable due to inappropriate security policies (Younis & Kifayat, 2013), lack of authentication and identity management (Megas et al., 2017), inadequate data security and privacy (Kaufman, 2009), lack of visibility (The Ponemon Institute, 2018), and insufficient security incidents management (Tweneboah-Koduah et al., 2017). For the *attack method*, the employee could use privilege escalation to generate false usernames and covertly steal digital files. In this case, the attack's impact was the theft of confidential information. The attack scenario is depicted in Fig. 8, showing the attack sequence where insiders can use privilege escalation to access a cloud storage/drive and exploit vulnerabilities such as a lack of authentication to access and steal confidential and proprietary information about products and processes.

Since the employee created false usernames and the system data was tampered with, anomalies could be observed in the data. Additionally, the employee might have logged in to the system from an external computer to download the data. Logging from an external network or an unauthorized device is a communication anomaly, and downloading digital files is an event-based anomaly. All these anomalies would be reflected in abnormal network traffic patterns and system log discrepancies. Hence, the knowledge of these anomalies can be utilized to design and implement several preventive and detective security measures. For example, attribute-based access control and multi-factor authentication (authentication schemes should be encrypted) for all logins are preventive countermeasures (Li et al., 2012). Distributed intrusion detection systems at both host and system levels are potential detective countermeasures.

#### *4.1.2 Attack on the process integrity*

The insider threat could also target hardware components (*attack target*), such as the human-machine interface, and physically tamper (*attack method*) with the production process. Furthermore, sensors used in production processes could be tampered with to provide fake or unreliable data, and the signal perceived by actuators could be altered so that the actuator implements wrong decisions (Trippel et al., 2017). In doing so, adversaries can exploit system vulnerabilities such as lack of visibility (The Ponemon Institute, 2018), insufficient authentication/authorization (Wells et al., 2014), poor identity management (Roman et al., 2011), and inappropriate usage of removable media devices (Clark et al., 2011) in hardware and IoT devices. Attacks on process integrity can lead to significant production disruptions, machine downtime, and manufacturing equipment sabotage. Alternations in the production sequence or process parameters will be reflected as anomalies in the process logs and/or in-situ process dynamics, providing an avenue to develop and deploy effective countermeasures against such attacks. Proper authorization, improved visibility and traceability, and access control are preventive countermeasures against this attack, while behavioral anomaly detection approaches at the hardware and process levels are detective controls. Potential threat attributes and countermeasures for this attack are also depicted in Fig. 8.

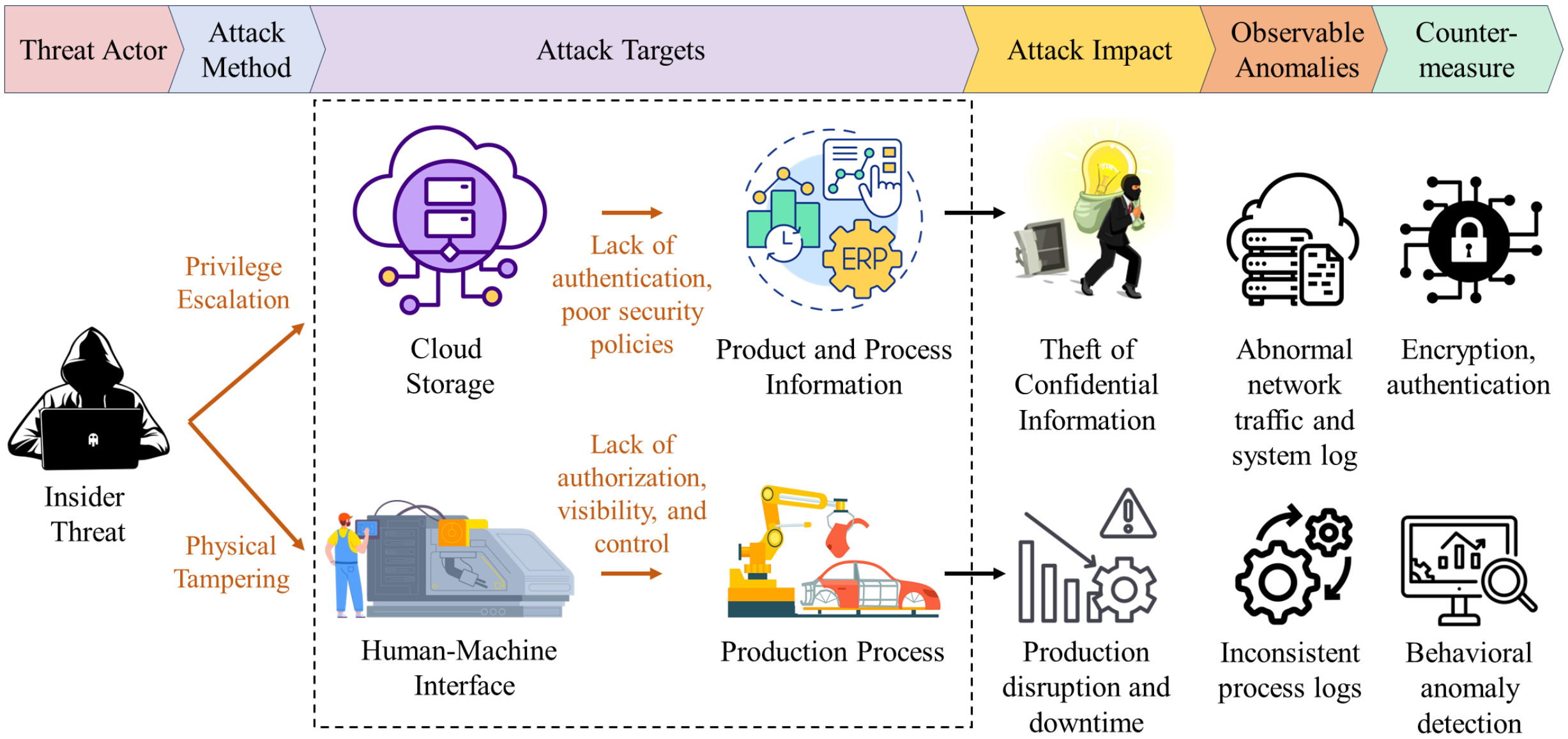

**Fig. 8**: Stealing of confidential information by an insider and a potential sabotage attack on the production process

### 4.2 Scenario 2: An attack on the geometric integrity of a product

Several researchers have demonstrated product-oriented cyber-physical attacks to reduce the product's functional performance, reliability, outgoing quality, and design intent. As *the attack target*, threat actors can tamper with CAD and tool path files, which are identified as sub-origins in this taxonomy. To access these files, they can target the network communication system (origin), exploiting its vulnerabilities, such as insecure network architecture (Fabro & Nelson, 2007), unreliable data transferring/sharing (Reaves & Morris, 2012), insufficient authentication/authorization (Moore et al., 2016), lack of encryption (Moore et al., 2016), insecure web interface (Costin et al., 2014), and usage of removable media devices (Clark et al., 2011). After gaining access, threat actors can tamper with the digital files (*attack method*) to insert voids at a part's stress concentration region. For example, threat actors can manipulate the geometry of the flank of the gear tooth near the dedendum circle, which encounters maximum stress during service. For additively manufactured gears, inserting voids in this region will dramatically reduce the part's strength, eventually causing the failure of the part. The attack discussed in scenario 2 is depicted in Fig. 9, showing the attack sequence of how the threat actor can use a web attack targeting the network communication system and digital files, exploit system vulnerabilities, and affect the geometric integrity of parts.

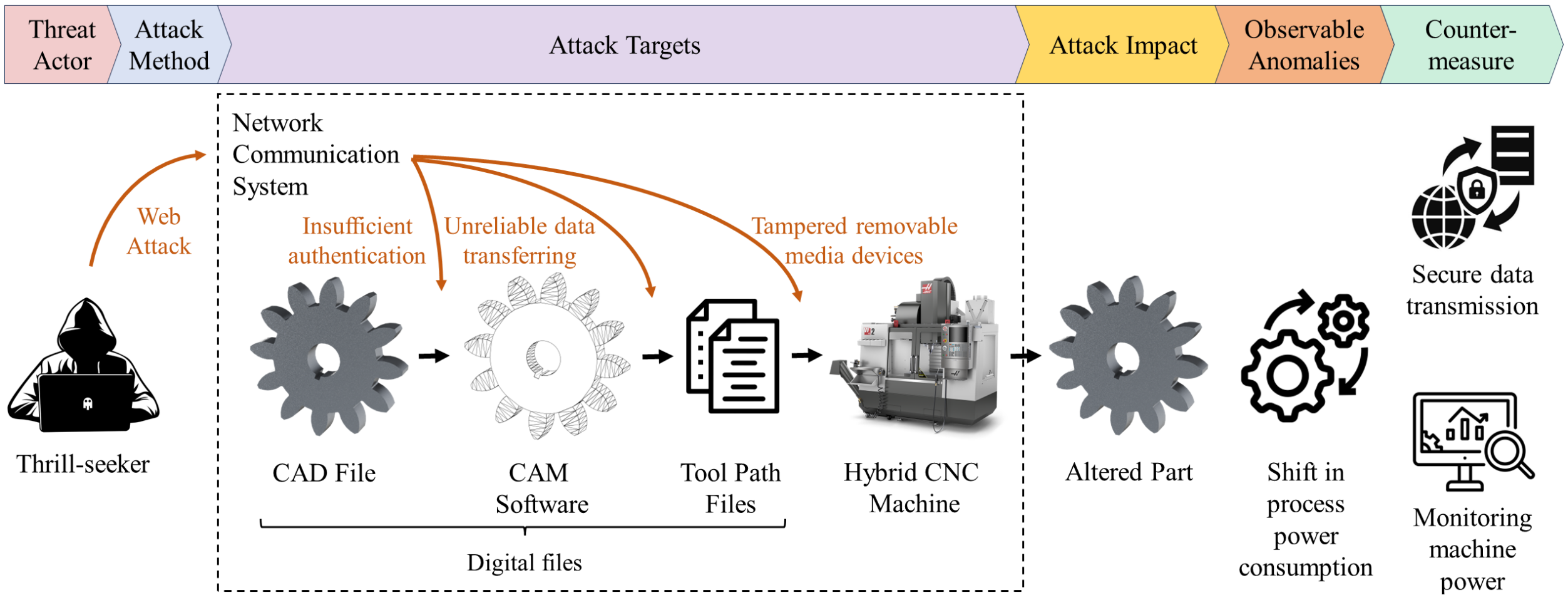

**Fig. 9**: Attack on the geometric integrity of a product leading to in-service failure of the part

Attacks on the part's geometric integrity will result in product-related impacts (*attack impact*). Informed by the proposed taxonomy, process and product anomalies can be observed due to the attack. The void might be designed by altering the wire feed rate and/or turning off the laser power in a Directed Energy Deposition (DED) process to skip a specific region or interrupt the material deposition. Depending on the attack strategy, the nominal print sequence or printing parameters will be changed while printing the tampered digital file. On the product side, there will be anomalies in the product's mechanical properties. These anomalies can be instantaneous or evolving based on the modification in the digital files. For example, briefly interrupting material deposition in each layer will result in an evolving anomaly across all print layers. On the other hand, deleting a specific G-code command, such as printing specific tracks, will result in an instantaneous anomaly in the process. Leveraging the understanding of potential anomalies in the process, appropriate countermeasures can be developed utilizing the process dynamics. For example, observing in-process variables such as acoustics and power consumption will detect addition, removal, and reordering in G-code commands (Belikovetsky et al., 2019). As a preventive countermeasure, secure data transmission across the network must be ensured (ICS-CERT, 2016). The network security system that monitors and controls traffic should be set up appropriately. Next-generation firewalls can be implemented to examine network traffic, filter out malicious websites, and prevent internet-based malware (IIROC & OCRCVM, 2015).

## 4.3 Reporting attack incidents

This section demonstrates how the proposed taxonomy can be utilized to understand, analyze, and describe cybersecurity threat attributes and potential countermeasures in real-world attack incidents and case studies. Table 3 shows how the different layers of the proposed taxonomy can aid in systematically describing the attack chain.

**Table 3**: Description of several real-world attacks and academic case studies using the proposed attack taxonomy

| Attack/Case study | Threat actor | Attack methods | Attack targets | | Attack impacts | Observable anomaly category | Countermeasures |
|---|---|---|---|---|---|---|---|
| | | | Origin | Sub-origin | | | |
| Stuxnet (*Stuxnet – Dangerous World*, 2018) | Nation-state actor | Zero-day attack | Communication (infected USB drive) | Commands (controller logic) | System-level impact | Process | Physics-based attack detection block in the controller (Giraldo et al., 2018) |
| Unitronics PLCs (Cyber, 2023) | Nation-state actor | Privilege escalation | Hardware (HMI) | Controller (PLC) | System-level impact | Process | Multi-factor authentication; access control (Stouffer et al., 2017) |
| German mill (R. M. Lee et al., 2014) | Cyber-criminal | Social engineering | Network communication system | Production system information | System-level impact | Process | Security awareness training; process monitoring (Stouffer et al., 2017) |
| FlashForge 3D Printer (McInerney, 2020) | Thrill-seeker | Web attack (spoofing) | Communication protocol | Machine firmware | System-level impact | Data | Access control; authentication (Stouffer et al., 2017) |
| Honda (Forbes, 2017) | Cyber-criminal | Denial of service, Web attack | Software, network communication | Controller | Operational impact | Data | Intrusion Prevention System (ICS-CERT, 2016) |
| (Wells et al., 2014) | Nation-state actor | Compromised software | Communication network | Tool path files | Product-related impact | Product (mechanical properties) | Security awareness and process monitoring (Stouffer et al., 2017) |
| (Chhetri et al., 2017) | Partner | Side-channel analysis | Manufacturing environment | Emissions | Theft of confidential information | - | Access control; noise injection (Mahesh et al., 2021; Stouffer et al., 2017) |
| (Belikovetsky et al., 2017) | Cyber-criminal | Phishing attack | Manufacturing environment | Design files | Product-related impact | Product (mechanical properties) | Acoustic emission monitoring (Belikovetsky et al., 2019) |
| (Shafae et al., 2019) | Cyber-criminal | Social engineering | Cloud storage, communication network | CAD files, tool path files | Product-related impact | Product (dimensional) | Spindle power monitoring (Rahman & Shafae, 2022) |

## 5 Taxonomy-Driven Cybersecurity Decision-Support in Smart Manufacturing

While the primary contribution of this work is the development of a comprehensive taxonomy for characterizing cybersecurity threats and countermeasures in smart manufacturing systems, the broader challenge of cybersecurity decision-making is the overarching goal. Towards bridging this gap, we present a taxonomy-guided decision-support framework that demonstrates how the structured knowledge encoded in the taxonomy can inform key decision-making processes in manufacturing security. The framework consists of five interrelated stages: (1) defining objectives and security context, (2) identifying manufacturing assets and associated threat attributes, (3) modeling cyber-physical risk, (4) performing risk assessment, and (5) selecting and deploying appropriate countermeasures to provide end-to-end guidance from threat identification to mitigation for practitioners and researchers. This practical, hands-on, step-by-step workflow illustrates how each taxonomy layer, coupled with other security tools, frameworks, and methodologies, can be leveraged to support security decision-making in manufacturing environments. The five stages of this framework are depicted in Fig. 10 and explained in detail in the following subsections.

## 5.1 Define objectives and security context

The decision-making process starts with establishing mission objectives, organizational priorities, and acceptable cybersecurity risk levels. This may include identifying compliance requirements such as NIST 800-171 and ISO/IEC 27001 (*ISO/IEC 27001 Information Security Management*, 2013; Ross et al., 2016), quality standards like ISO 9001 (*ISO 9001:2015*, 2021), and the potential attack consequences such as production downtime, safety hazards, or delivery delays—which will vary depending on the type of industry and product line. For example, a defense contractor handling classified information may prioritize data integrity and access control, whereas a general sheet metal manufacturer may focus on minimizing unplanned downtime. These contextual inputs determine which *threat actors* are most relevant to a particular manufacturing environment and how attack impacts are interpreted. In this preliminary stage, organizations can adopt existing frameworks, such as the Cybersecurity Maturity Model Certification (CMMC), to align their cybersecurity goals with regulatory requirements, especially in defense-related manufacturing sectors (*CMMC Program*, 2024). Such frameworks can assist manufacturers in determining their cybersecurity readiness levels and improving their security posture in proportion to the sensitivity of the data and operations they manage.

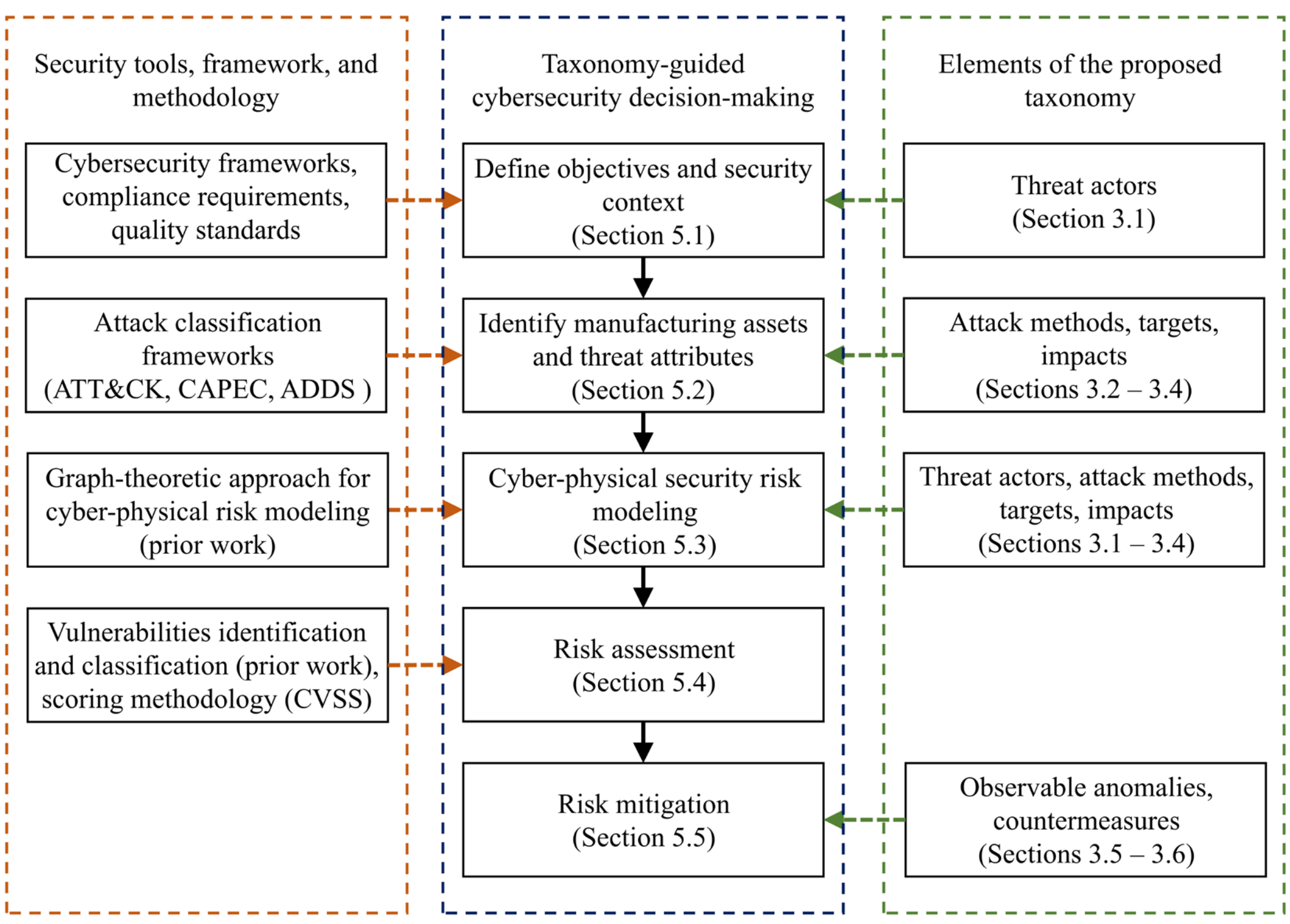

**Fig. 10**: Taxonomy-guided cybersecurity decision-support in smart manufacturing systems

## 5.2 Identify manufacturing assets and threat attributes

This task involves systematically identifying and associating critical manufacturing assets with relevant threat attributes, forming the basis for systematic threat modeling and risk assessment. First, a comprehensive asset inventory should be developed that includes physical components such as machinery, controllers, sensors, metrology systems, and human-machine interfaces; digital assets such as software, firmware, and cloud-connected applications; and network communication infrastructures, including data flows, protocols, and network interfaces. End-user devices, including employee-owned devices linked to the manufacturing network, must also be considered, as they frequently introduce unmonitored access points into the network system. Asset inventories must be regularly updated to reflect changing configurations, third-party integrations, and user access.

Second, each asset's criticality is determined by its role in ensuring product quality, operational continuity, and regulatory compliance. Critical assets are then assigned to the taxonomy's *attack target*, *attack method*, and *attack impact* layers. The attack target layer identifies system components that are likely to be compromised, and the attack method layer specifies the techniques or tactics that will be used to exploit the asset. The impact layer characterizes the system-level consequences if an asset is compromised, such as production halts and quality degradation. At this stage, organizations may utilize cybersecurity tools, frameworks, and repositories such as the MITRE ATT&CK framework, Attack Design Designation System (ADDS), and CAPEC to identify relevant attack vectors, techniques, and procedures (*CAPEC*, 2023; *MITRE*, 2022; Shafae et al., 2019).

Assigning specific threat attributes to individual assets depends on contextual knowledge of the system and typically requires input from industry practitioners or domain experts familiar with the manufacturing environment. For example, a network-accessible Programmable Logic Controller (PLC) may be identified as an attack target due to its critical role in executing machine instructions. If compromised, the impact could include executing unauthorized commands that can lead to part defects or machine damage. In this case, potential attack methods may consist of command injection, particularly without strong authentication protocols. In contrast, sensors on the machine shop floor may be classified as targets with consequences, including false data capture or process disruption. It may be exploited using physical attack methods such as physical tampering (e.g., recalibration) or signal spoofing, primarily if no tamper-detection features exist. Additionally, threat actors often adopt specific types of attack methods that reflect their strategic intent and technical capabilities. For example, the cybercriminal group XENOTIME, responsible for the 2017 attack on an oil and gas facility in Saudi Arabia, has been identified as having capabilities designed to compromise Industrial Control Systems (ICSs) (*Dragos*, 2020). Groups like XENOTIME are known to employ signature tactics that target manufacturing, energy, and defense sectors. Identifying and characterizing threat actors and their signature attack methods will require a combination of technical expertise and access to threat intelligence (Rahman et al., 2023).

## 5.3 Cyber-physical security risk modeling

Graph-theoretic modeling approaches are particularly well-suited for systematically representing and analyzing cybersecurity risks, as they effectively capture complex relationships among diverse threat attributes and attack pathways. Leveraging this suitability, we previously developed a graph-theoretic framework capable of concurrently modeling cybersecurity threats and quantitatively assessing associated risks (Rahman et al., 2024). The comprehensive taxonomy presented in Section 3 of this study complements and strengthens this graph-theoretic approach by providing a structured basis for generating well-defined and systematic attack graphs, overcoming limitations inherent in traditional ad-hoc or experience-based methods. In this section and the following, we briefly summarize and integrate our earlier risk modeling and assessment framework into the proposed taxonomy-guided decision-support framework, highlighting its enhanced effectiveness within this structured context. (Rahman et al., 2024)The attack graph formalism can help model the interrelation between attack methods, interdependent attack targets, and potential impacts on manufacturing assets, as depicted in Fig. 11 (left). By incorporating different attack attributes in the risk model, attack graphs allow concurrent modeling and analysis of various cybersecurity threats comprising varying attack methods, targets, vulnerabilities, and impacts. It can be used to proactively explore how varying sequences of adversarial actions comprising different attack paths through the system can affect diverse cyber and/or physical assets to realize the intended attack consequence.

The attack graph is a triplet: $AG = (V, E, W)$. $V \subset \{1, \dots, N\}$ is the set of $N$ vertices satisfying $V = \{V_{AV}\} \cup \{V_L\} \cup \{V_C\}$, where $V_{AV}$ are the vertices representing attack methods threat actors use to launch an attack. Potential attack targets are represented using $V_L$, and $V_C$ are the vertices depicting attack impacts. To facilitate the notation, nodes are denoted by lower-case $v$ and a subscript denoting its label. $v_{av_i} \in V_{AV}$ denotes the $i^{\text{th}}$ attack vector, $i \subset \{1, \dots, N_{AV}\}$; $v_{l_j} \in V_L$ denotes the $j^{\text{th}}$ attack target, $j \subset \{1, \dots, N_L\}$; $v_{c_k} \in V_C$ denotes the $k^{\text{th}}$ attack consequence, $k \subset \{1, \dots, N_C\}$. $E \subset V \times V$ is the set of directed edges representing connections between different vertices. $W \subset R^N \times R^N$ represents the weight of different edges. The directed edges in the graph can portray the information (cyber) and material (physical) flow (i.e., cyber-physical interconnectivity among different vertices). The weight of the edge is analogous to the distance between the two connected vertices, representing how difficult it is to reach and compromise one vertex from another in the given direction.

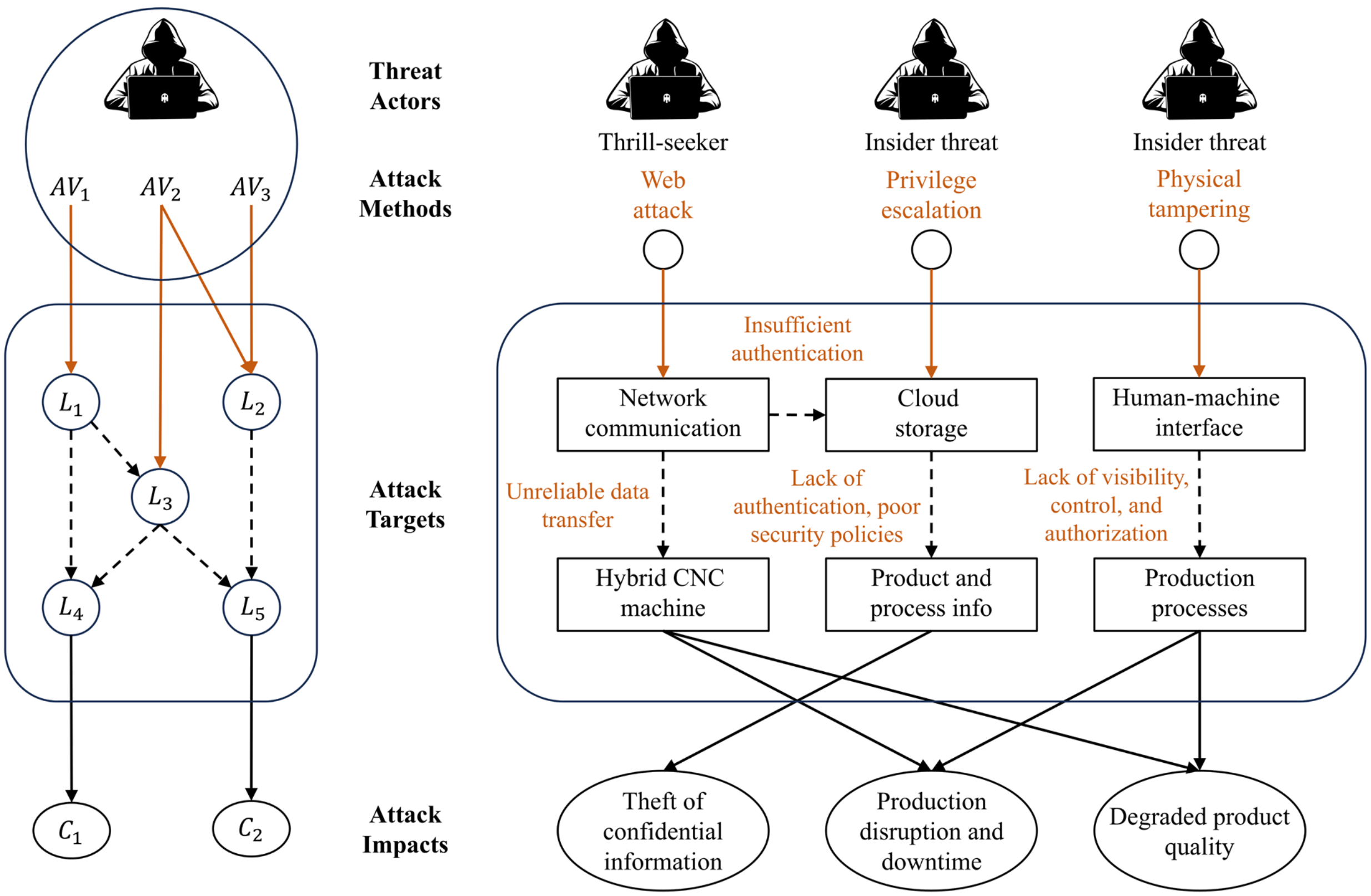

**Fig. 11**: Graph-theoretic model of cyberattacks on manufacturing systems (left), sample attack graphs for the examples presented in Sections 4.1 and 4.2 (right)

Organizations can use the proposed taxonomical classification of attack attributes along with security tools like MITRE ATT&CK framework (*MITRE*, 2022) and ADDS (Shafae et al., 2019) to populate nodes in the attack graph. Next, they can construct graph edges using expert opinion, domain knowledge of system interdependencies, and historical data on attack feasibility. Sample attack graphs for the attacks described in Sections 4.1 and 4.2 are presented in Fig. 11 (right). To illustrate, an insider threat can target hardware components, such as the human-machine interface, and physically tamper with the production process. Production processes can be further classified as machining (subtractive), additive, joining, and assembly processes. Machining processes can be classified into turning, drilling, milling, boring, laser cutting, etc. Similarly, additive manufacturing, joining, and assembly processes can be divided into sub-categories. With the necessary decomposition for an organization, all these individual processes can be considered separate vertices in the more granular attack graph. Such detailed decomposition can be helpful given the varying nature of attack targets and impacts depending on the production process type. This will aid in modeling the spread of a multi-stage attack to manipulate product quality at various manufacturing stages. Similarly, all other attack attributes can be expanded to fill the attack graph's vertices, and appropriate edges can be added to represent their interconnections.

## 5.4 Risk assessment

The cybersecurity threat event is a sequence of activities from attack launches to inflicting damage to the system modeled as directed paths within an attack graph. A fundamental step in risk assessment is quantifying the likelihood of an adversary successfully executing such a sequence. Cybersecurity risk assessment involves quantifying (1) the probability of compromising individual assets, (2) the likelihood of attack propagation across interdependent system nodes, and (3) the total risk associated with a given attack path, which is briefly explained in the following subsections.

#### *5.4.1 Probability of compromising an asset*

The probability of compromising a specific attack target $v_{l_j} \in V_L$ using attack vector $v_{av_i} \in V_{AV}$ is determined as: $P_{AV_i,L_j} = AV \times AC \times PR \times UI \times \frac{1}{RL}$. Here, $P_{AV_i,L_j}$ denotes the probability of $i^{\text{th}}$ attack vector compromising the $j^{\text{th}}$ attack target, which is a function of several metrics that collectively determine an asset's vulnerability. $AV, AC, PR, UI$, and $RL$ represent the attack vector, attack complexity, privilege required, user interaction, and remediation level metrics, respectively. Organizations can define the specific scoring system for those metrics as well as adopt an external scoring system such as NIST's Common Vulnerability Scoring System (CVSS) (*Common Vulnerability Scoring System*, 2023). Here, $1/P_{AV_i,L_j}$ denotes the weight of the edge from any vertex $v_{av_i} \in V_{AV}$ to vertex $v_{l_j} \in V_L$, i.e., $W_{av_i,l_j}$. The lower the value of $W_{av_i,l_j}$, the easier it would be for threat actors to compromise vertex $v_{l_j}$ from vertex $v_{av_i}$.

#### *5.4.2 Likelihood of attack propagation*

The likelihood of attack propagation from one vertex to another along the attack path can be modeled as: $T_{m,m+1} = Pr(x_{m+1}|x_m)$, where $m \in V_L, V_C$. Here, $x_m$ represents that vertex $m$ has been compromised, and $T_{m,m+1}$ represents the fault propagation probability from $m^{\text{th}}$ vertex to $(m+1)^{\text{th}}$ vertex in the attack path. $T_{m,m+1}$ can be considered deterministic for simplifying the calculation, or the transition probability can be dynamically updated based on gained knowledge on adversarial strategies for dynamic risk assessment. For example, the probability can be updated based on observed sequences in previous attacks, process mining and alert correlation, and post-attack forensics.

#### *5.4.3 Risk estimation*

Total risks of potential attack paths can be calculated from the probabilities of compromising individual attack targets, attack propagation, and expected losses associated with the consequence in the attack graph. The risk of an attack path is calculated as:

$$R = P_{AV_i,L_j} \left( \prod_{m=1}^{n-1} T_{v_\lambda^m, v_\lambda^{m+1}} \right) T_{v_\lambda^n, v_{c_k}} C_k$$

Here, $P_{AV_i,L_j}$ denotes the probability of compromising the attack target $v_{l_j}$ using attack vector $v_{av_i}$. $T_{v_\lambda^m, v_\lambda^{m+1}}$ represents attack propagation from the vertex $v_\lambda^m$ to vertex $v_\lambda^{m+1}$ in the attack path, $n$ is the total number of attack targets compromised during the threat event. $\{v_\lambda^\beta\}_{\beta \subset \{1,\ldots,n\}}$ represents the ordered set of compromised assets. $T_{v_\lambda^n, v_{c_k}}$ is the transition probability from the last vertex of the $\{v_\lambda^\beta\}_{\beta \subset \{1,\ldots,n\}}$ subsequence to attack consequence $v_{c_k}$. $C_k$ is the impact (e.g., monetary loss, safety risk, product failure) tied to the attack's outcome. Manufacturers will specify $v_{av_i}$ and $v_{c_k}$, based on the threat attributes applicable to their manufacturing ecosystem. $C_k$ can be quantified as (1) theft of confidential information, (2) compromised system integrity, and (3) system unavailability. In the manufacturing domain, more specific business consequences such as (1) lost sales, (2) increased production waste, (3) recovery costs from sabotage or system damage, (4) cost of operational downtime, (5) repair costs from machine breakdown, (6) personnel safety risk, and (7) product-related damages (including reputation damage from poor product quality and reliability), can be estimated. The quantitative model will estimate the cybersecurity risk associated with various attack paths. The probability of attack likelihood and attack propagation used in the assessment must be defined by the organization using a combination of system-specific knowledge, identified system vulnerabilities, threat modeling insights, and scoring methodologies like the CVSS to quantify the likelihood and complexity of exploiting individual assets (*Common Vulnerability Scoring System*, 2023). To support this process, our prior work introduces a systematic approach for characterizing and identifying vulnerabilities across the manufacturing cyberspace, human element, post-production inspection systems, production process monitoring, and organizational policies and procedures (Rahman & Shafae, 2025).

Therefore, manufacturing-specific taxonomical classification of threat attributes can assist practitioners in creating comprehensive and generalizable cyber-physical attack graphs for modeling the cascading attack impact that leads to specific outcomes. Graph search algorithms can be used to computationally analyze the attack graph, allowing organizations to explore all possible attack scenarios or prioritize the most likely/threatening attack path. For example, the Depth-First Search (DFS) algorithm can be used to identify all attack paths that originate from a given attack vector, whereas Dijkstra's algorithm can be used to identify the shortest attack path from the adversary's perspective—defined as the path with the lowest cumulative weight (i.e., the highest probability of success). The shortest attack path poses the maximum risk for manufacturers and highlights the most critical manufacturing assets that need prioritized security controls and defense measures.(Rahman et al., 2024)

### 5.5 Risk mitigation

Once the critical assets are identified through taxonomy-driven graph-theoretic risk modeling and assessment, risk mitigation involves developing and deploying proactive and reactive defense strategies to reduce exposure and limit the impact of cyberattacks. Proactive defenses are implemented to prevent the success of attacks by hardening systems and eliminating known vulnerabilities, while reactive defenses focus on detecting, diagnosing, and containing attacks once they occur. The selection and prioritization of countermeasures are informed by risk assessment and guided by the proposed taxonomy's *observable anomalies* and *countermeasures* layers. Risk mitigation strategies can focus on three key objectives: (1) mitigating vulnerabilities in critical manufacturing assets to reduce the likelihood of exploitation, (2) designing and deploying anomaly-informed defenses to support early attack detection and diagnosis, and (3) applying graph-theoretic tools (e.g., minimum cut, attack path analysis) to contain and limit the spread of attacks once detected.

The *observable anomalies* layer characterizes how cyber-physical attacks frequently manifest as deviations in system behavior across three domains: data, process, and product. These observable anomalies can provide early indicators of malicious activity, making them critical for developing effective detection strategies. For example, malicious alterations of product design files will introduce geometric deviations in the final part and/or change machining behavior. Such manifestations can be instantaneous, evolve over time, or result from the accumulation of subtle anomalies at multiple stages of production. Identifying the type and location of these anomalies allows manufacturers to implement domain-specific, context-aware detection mechanisms. The proposed taxonomy also provides a structured classification of *countermeasures* across three dimensions: activity level, deployment location, and response strategy, which can help determine where countermeasures should be deployed and what function they should perform. For instance, watermarking serves a preventive and diagnostic function at the design stage; physics-informed process monitoring offers real-time alerting during production; and IDS or anomaly correlation tools provide alert and forensic capabilities distributed across the network. The *observable anomalies* and *countermeasures* layers offer complementary insights into how attacks are realized and how they can be countered across the manufacturing system. Examples of potential countermeasures are presented in Section 3.6.2, different observable anomalies and corresponding countermeasures are discussed in Section 4, and a more detailed discussion on vulnerability mitigation strategies is provided in our prior work (Rahman & Shafae, 2025).

## 6 Conclusion

This study develops a comprehensive manufacturing-specific attack taxonomy composed of six layers to systematically characterize and classify cybersecurity threat attributes and potential countermeasures. It introduces structured classifications for threat actors and their intent, categorizes attack methods and targets, and expands the existing high-level categorization of attack impacts by incorporating system-level and operational consequences specific to manufacturing environments. The comprehensive taxonomy also characterizes system behavioral deviations caused by attacks, distinguishing them from natural process variations and classifying them by severity and impact. Additionally, this paper presents a classification structure for countermeasures, provides representative examples, and demonstrates how they align with the proposed taxonomical framework. Unlike prior efforts, it enables a structured, end-to-end characterization of cyber-physical attacks, from adversarial motivation to observable system deviations and appropriate defense strategies. The proposed taxonomy's applicability is supported through two illustrative attack scenarios, several real-world cyber-physical attack incidents, and academic case studies, illustrating how the taxonomy aids in systematically documenting and analyzing manufacturing cybersecurity threats. Building on this foundation, a taxonomy-guided decision-support framework is introduced to demonstrate how the proposed taxonomy complements other cybersecurity tools, frameworks, and methodologies to support integrated, risk-informed security decision-support in manufacturing environments, from the selection of threat attributes and risk modeling to risk assessment, and appropriate countermeasure development and deployment.

Future work will augment this work by including novel attack methods, their impacts, and an exhaustive analysis of all the potential preventive and detective countermeasures that can be adopted to defend the system from external or internal attacks. Another research direction involves developing an ontology to incorporate the proliferation of knowledge on manufacturing-specific threat attributes and countermeasures, which will require a thorough conceptual analysis of different attributes of cyberattacks and their interconnections. Additionally, integrating the proposed taxonomy into widely adopted cybersecurity audit frameworks, such as those developed by NIST, could enhance its practical value, particularly for Small and Medium-sized Enterprises (SMEs) with limited security resources. Future efforts will also explore the creation of a structured audit questionnaire based on the proposed taxonomy to support SMEs in assessing cyber-physical security risks and selecting cost-effective and context-appropriate countermeasures. This would provide a coordinated, resource-efficient approach for improving cybersecurity posture across digital manufacturing environments.

The proposed taxonomy provides a holistic risk landscape by systematically classifying threat attributes and linking them to system anomalies and countermeasures. It also creates a practical foundation for manufacturing security awareness and designing and operating resilient smart manufacturing systems. This paper provides the first end-to-end approach for manufacturing cybersecurity practitioners and researchers, from threat identification to mitigation in manufacturing systems.

## Statements and Declarations

### Competing Interest

The authors declare that they have no known competing financial interests or personal relationships that could have appeared to influence the work reported in this paper.

### Acknowledgment

This work has been partially funded through the Technology and Research Initiative Fund (TRIF) under the National Security Systems Initiative, funded under Proposition 301, the Arizona Sales Tax for Education Act. Additionally, this work was supported by the National Science Foundation under Grant No. 2119654 & 2519581. Any opinions, findings, conclusions, or recommendations expressed in this material are those of the author(s) and do not necessarily reflect the views of the funding agencies.

## References

Ahmed, M., & Pathan, A.-S. K. (2020). False data injection attack (FDIA): an overview and new metrics for fair evaluation of its countermeasure. *Complex Adaptive Systems Modeling*, *8*(1), 1–14. https://doi.org/https://doi.org/10.1186/s40294-020-00070-w

Belikovetsky, S., Solewicz, Y. A., Yampolskiy, M., Toh, J., & Elovici, Y. (2019). Digital Audio Signature for 3D Printing Integrity. *IEEE Transactions on Information Forensics and Security*, *14*(5), 1127–1141. https://doi.org/10.1109/TIFS.2018.2851584

Belikovetsky, S., Yampolskiy, M., Toh, J., Gatlin, J., & Elovici, Y. (2017). Dr0wned – Cyber-physical attack with additive manufacturing. *11th USENIX Workshop on Offensive Technologies, WOOT 2017, Co-Located with USENIX Security 2017*.

Bhushan, B., Sahoo, G., & Rai, A. K. (2017). Man-in-the-middle attack in wireless and computer networking—A review. *2017 3rd International Conference on Advances in Computing, Communication & Automation (ICACCA)(Fall)*, 1–6.

Bilge, L., & Dumitras, T. (2012). Before We Knew It: An Empirical Study of Zero-Day Attacks In The Real World Categories and Subject Descriptors. *Proceedings of the 2012 ACM Conference on Computer and Communications Security - CCS '12*, 833–844. https://doi.org/10.1145/2382196.2382284

Blank, R. M., & Gallagher, P. D. (2012). *Guide for conducting risk assessments*. https://doi.org/10.6028/NIST.SP.800-30r1

*Bloomberg*. (2021). https://www.bloomberg.com/news/articles/2021-06-04/hackers-breached-colonial-pipeline-using-compromised-password

*Canadian Centre for Cyber Security*. (2021). https://cyber.gc.ca/en/guidance/cyber-threat-and-cyber-threat-actors

*CAPEC*. (2023). https://capec.mitre.org/

Chen, F., Luo, Y., Tsoutsos, N. G., Maniatakos, M., Shahin, K., & Gupta, N. (2019). Embedding tracking codes in additive manufactured parts for product authentication. *Advanced Engineering Materials*, *21*(4), 1800495. https://doi.org/https://doi.org/10.1002/adem.201800495

Chhetri, S. R., Canedo, A., & Faruque, M. A. Al. (2017). Confidentiality breach through acoustic side-channel in cyber-physical additive manufacturing systems. *ACM Transactions on Cyber-Physical Systems*, *2*(1), 1–25. https://doi.org/https://doi.org/10.1145/3078622

Clark, J., Leblanc, S., & Knight, S. (2011). Risks associated with USB Hardware Trojan devices used by insiders. *2011 IEEE International Systems Conference*, 201–208. https://doi.org/10.1109/SYSCON.2011.5929130

*CMMC Program*. (2024, October 15). https://www.federalregister.gov/documents/2024/10/15/2024-22905/cybersecurity-maturity-model-certification-cmmc-program

*CNBC*. (2018, June 18). https://www.cnbc.com/2018/06/18/elon-musk-email-employee-conducted-extensive-and-damaging-sabotage.html

*CNN Business*. (2022). https://www.cnn.com/2022/03/01/business/toyota-japan-cyberattack-production-restarts-intl-hnk/index.html

*Common Vulnerability Scoring System*. (2023). https://www.first.org/cvss/specification-document

Costin, A., Zaddach, J., Francillon, A., Balzarotti, D., Sophia, E., & France, A. (2014). A Large-Scale Analysis of the Security of Embedded Firmwares. *23rd {USENIX} Security Symposium ({USENIX} Security 14)*, 95–110. http://firmware.re

*CVE-2020-14750*. (2020). https://www.oracle.com/security-alerts/alert-cve-2020-14750.html

Cyber, I. (2023). *US, Israeli security agencies warn of Iranian IRGC executing malicious cyber activity using OT devices*. https://industrialcyber.co/cisa/us-israeli-security-agencies-warn-of-iranian-irgc-executing-malicious-cyber-activity-using-ot-devices/

*Cybersecurity Magazine*. (2023). https://www.uscybersecurity.net/script-kiddie/

*Deloitte*. (2022). https://www2.deloitte.com/ca/en/pages/risk/articles/the-changing-faces-of-cybersecurity.html

*Dragos*. (2020). https://www.dragos.com/resource/manufacturing-threat-perspective/

El Abbadi, R., & Jamouli, H. (2021). Takagi–Sugeno Fuzzy Control for a Nonlinear Networked System Exposed to a Replay Attack. *Mathematical Problems in Engineering*, *2021*. https://doi.org/https://doi.org/10.1155/2021/6618105

Elhabashy, A. E., Wells, L. J., Camelio, J. A., & Woodall, W. H. (2019). A cyber-physical attack taxonomy for production systems: a quality control perspective. *Journal of Intelligent Manufacturing*, *30*(6), 2489–2504. https://doi.org/10.1007/s10845-018-1408-9

Esmaeilian, B., Behdad, S., & Wang, B. (2016). The evolution and future of manufacturing: A review. *Journal of Manufacturing Systems*, *39*, 79–100. https://doi.org/https://doi.org/10.1016/j.jmsy.2016.03.001

Fabro, M., & Nelson, T. (2007). Control systems cyber security: defense in depth strategies. In *Idaho National Laboratory, USA*. https://inldigitallibrary.inl.gov/sites/sti/sti/3867714.pdf

FireEye. (2023). *What is a Zero-Day Exploit? | FireEye*. https://us.norton.com/blog/emerging-threats/how-do-zero-day-vulnerabilities-work

Forbes. (2017). *Cyber Attack At Honda Stops Production After WannaCry Worm Strikes*. https://www.forbes.com/sites/peterlyon/2017/06/22/cyber-attack-at-honda-stops-production-after-wannacry-worm-strikes/#12919d15e2b7

*Fortune*. (2022). https://fortune.com/2022/03/23/nestle-anonymous-leak-hack-russia-business-kitkat-nesquik/

*Framework for Improving Critical Infrastructure Cybersecurity, Version 1.1*. (2018). https://doi.org/10.6028/NIST.CSWP.04162018

Giraldo, J., Urbina, D., Cardenas, A., Valente, J., Faisal, M., Ruths, J., Tippenhauer, N. O., Candell, R., Tippen-Hauer, N. O., & Sandberg, H. (2018). A Survey of Physics-Based Attack Detection in Cyber-Physical Systems. *ACM Comput. Surv.*, *51*(4), 1–32. https://doi.org/10.1145/3203245

Graves, L. M. G., King, W., Carrion, P., Shao, S., Shamsaei, N., & Yampolskiy, M. (2021). Sabotaging metal additive manufacturing: Powder delivery system manipulation and material-dependent effects. *Additive Manufacturing*, 102029. https://doi.org/https://doi.org/10.1016/j.addma.2021.102029

Howard, J. D., & Longstaff, T. A. (1998). A common language for computer security incidents. In *Sandia National Laboratories* (Vol. 10). https://doi.org/https://doi.org/10.2172/751004

IBM. (2021). *IBM Security X-Force Threat Intelligence Index*. https://www.ibm.com/security/data-breach/threat-intelligence

*IBM*. (2025). https://www.ibm.com/thought-leadership/institute-business-value/en-us/report/2025-threat-intelligence-index

ICS-CERT. (2016). *Improving Industrial Control System Cybersecurity with Defense-in-Depth Strategies*. https://www.cisa.gov/sites/default/files/recommended_practices/NCCIC_ICS-CERT_Defense_in_Depth_2016_S508C.pdf

IIROC, & OCRCVM. (2015). *Cybersecurity Best Practices Guide For IIROC Dealer Members*. https://mcmillan.ca/wp-content/uploads/2016/07/CybersecurityBestPracticesGuide_en.pd_.pdf

*ISO/IEC 27001 Information security management*. (2013). https://www.iso.org/isoiec-27001-information-security.html

*ISO 9001:2015*. (2021). https://www.iso.org/standard/62085.html

Kaufman, L. M. (2009). Data security in the world of cloud computing. *IEEE Security & Privacy*, *7*(4), 61–64. https://doi.org/https://doi.org/10.1109/MSP.2009.87

Khraisat, A., Gondal, I., Vamplew, P., & Kamruzzaman, J. (2019). Survey of intrusion detection systems: techniques, datasets and challenges. *Cybersecurity*, *2*(1), 1–22. https://doi.org/https://doi.org/10.1186/s42400-019-0038-7

Lee, H.-T., Kim, D., Park, M., & Cho, S. (2016). Protecting data on android platform against privilege escalation attack. *International Journal of Computer Mathematics*, *93*(2), 401–414.

Lee, R. M., Assante, M. J., & Conway, T. (2014). German steel mill cyber attack. In *Industrial Control Systems* (Vol. 30). https://media.kasperskycontenthub.com/wp-content/uploads/sites/20/2015/04/20000144/ICS-CPPE-case-Study-2-German-Steelworks_Facility.pdf

Li, X., Liang, X., Lu, R., Shen, X., Lin, X., & Zhu, H. (2012). Securing smart grid: cyber attacks, countermeasures, and challenges. *IEEE Communications Magazine*, *50*(8), 38–45. https://doi.org/10.1109/MCOM.2012.6257525

Lin, Y.-Z., Shao, S., Rahman, M. H., Shafae, M., & Satam, P. (2023). DT4I4-Secure: Digital Twin Framework for Industry 4.0 Systems Security. *2023 IEEE 14th Annual Ubiquitous Computing, Electronics & Mobile Communication Conference (UEMCON)*, 200–209. https://doi.org/https://doi.org/10.1109/UEMCON59035.2023.10316090

Lopez, F., Saez, M., Shao, Y., Balta, E. C., Moyne, J., Mao, Z. M., Barton, K., & Tilbury, D. (2017). Categorization of anomalies in smart manufacturing systems to support the selection of detection mechanisms. *IEEE Robotics and Automation Letters*, *2*(4), 1885–1892. https://doi.org/https://doi.org/10.1109/LRA.2017.2714135

Maes, R., & Verbauwhede, I. (2010). Physically unclonable functions: A study on the state of the art and future research directions. In *Towards Hardware-Intrinsic Security* (pp. 3–37). Springer. https://doi.org/https://doi.org/10.1007/978-3-642-14452-3_1

Mahesh, P., Tiwari, A., Jin, C., Kumar, P. R., Reddy, A. L. N., Bukkapatanam, S. T. S., Gupta, N., & Karri, R. (2021). A Survey of Cybersecurity of Digital Manufacturing. *Proceedings of the IEEE*, *109*(4), 495–516. https://doi.org/10.1109/JPROC.2020.3032074

McInerney, D. (2020). *With IoT, Common Devices Pose New Threats*. Coalfire. https://www.coalfire.com/the-coalfire-blog/april-2020/with-iot-common-devices-pose-new-threats

Megas, K., Piccarreta, B., & O'Rourke, D. G. (2017). *Internet of things (IoT) cybersecurity colloquium: a NIST workshop proceedings*. https://doi.org/10.6028/NIST.IR.8201

Mehdi, N., & Starly, B. (2020). Witness Box Protocol: Automatic machine identification and authentication in industry 4.0. *Computers in Industry*, *123*, 103340.

*MITRE*. (2022). https://attack.mitre.org/groups/

Moore, S., Armstrong, P., McDonald, T., & Yampolskiy, M. (2016). Vulnerability analysis of desktop 3D printer software. *Proceedings - 2016 Resilience Week, RWS 2016*, 46–51. https://doi.org/10.1109/RWEEK.2016.7573305

NIST. (2025). *CSRC*. https://csrc.nist.gov/glossary/term/tampering

Pan, Y., White, J., Schmidt, D. C., Elhabashy, A., Sturm, L., Camelio, J., & Williams, C. (2017). Taxonomies for Reasoning About Cyber-physical Attacks in IoT-based Manufacturing Systems. *International Journal of Interactive Multimedia & Artificial Intelligence*, *4*(3), 45–54. https://doi.org/10.9781/ijimai.2017.437

Rahman, M. H. (2024). *Secure Cyber-Physical Manufacturing Systems (Secure-CyPhyMan): Advanced Methods for Manufacturing Cybersecurity Threat Characterization, Risk Modeling and Assessment, and Resilient Attack Detection and Prevention*. The University of Arizona.

Rahman, M. H., Hamedani, E. Y., Son, Y.-J., & Shafae, M. (2024). Taxonomy-Driven Graph-Theoretic Framework for Manufacturing Cybersecurity Risk Modeling and Assessment. *Journal of Computing and Information Science in Engineering*, *24*(7), 071003. https://doi.org/10.1115/1.4063729

Rahman, M. H., & Shafae, M. (2022). Physics-based detection of cyber-attacks in manufacturing systems: a machining case study. *Journal of Manufacturing Systems*, *64*, 676–683. https://doi.org/10.1016/j.jmsy.2022.04.012

Rahman, M. H., & Shafae, M. (2025). Cyber-Physical Security Vulnerabilities Identification and Classification in Smart Manufacturing: A Defense-in-Depth Driven Framework and Taxonomy. *Journal of Computing and Information Science in Engineering*, *25*(9), 091005. https://doi.org/https://doi.org/10.1115/1.4068844

Rahman, M. H., Wuest, T., & Shafae, M. (2023). Manufacturing cybersecurity threat attributes and countermeasures: Review, meta-taxonomy, and use cases of cyberattack taxonomies. *Journal of Manufacturing Systems*, *68*, 196–208. https://doi.org/https://doi.org/10.1016/j.jmsy.2023.03.009

Reaves, B., & Morris, T. (2012). Analysis and mitigation of vulnerabilities in short-range wireless communications for industrial control systems. *International Journal of Critical Infrastructure Protection*, *5*(3–4), 154–174. https://doi.org/10.1016/j.ijcip.2012.10.001

Roman, R., Najera, P., & Lopez, J. (2011). Securing the internet of things. *Computer*, *9*, 51–58. https://doi.org/https://doi.org/10.1109/MC.2011.291

Ross, R., Viscuso, P., Guissanie, G., Dempsey, K., & Riddle, M. (2016). NIST Special Publication 800-171 Revision 1: Protecting controlled unclassified information in nonfederal systems and organizations. *NIST Special Publication 800-Series*, 21. https://doi.org/10.6028/NIST.SP.800-171r1

Sailio, M., Latvala, O. M., & Szanto, A. (2020). Cyber threat actors for the factory of the future. *Applied Sciences (Switzerland)*, *10*(12). https://doi.org/10.3390/app10124334

Sánchez-Fernández, A., Baldán, F. J., Sainz-Palmero, G. I., Benítez, J. M., & Fuente, M. J. (2018). Fault detection based on time series modeling and multivariate statistical process control. *Chemometrics and Intelligent Laboratory Systems*, *182*, 57–69. https://doi.org/https://doi.org/10.1016/j.chemolab.2018.08.003

Sandborn, M., Olea, C., White, J., Williams, C., Tarazaga, P. A., Sturm, L., Albakri, M., & Tenney, C. (2021). Towards secure cyber-physical information association for parts. *Journal of Manufacturing Systems*, *59*, 27–41. https://doi.org/https://doi.org/10.1016/j.jmsy.2021.01.003

Shafae, M. S., Wells, L. J., & Purdy, G. T. (2019). Defending against product-oriented cyber-physical attacks on machining systems. *The International Journal of Advanced Manufacturing Technology*, 1–21. https://doi.org/10.1007/s00170-019-03805-z

Shewhart, W. A. (1930). Economic quality control of manufactured product 1. *Bell System Technical Journal*, *9*(2), 364–389. https://doi.org/https://doi.org/10.1002/j.1538-7305.1930.tb00373.x

Simmons, C., Shiva, S., Bedi, H., & Dasgupta, D. (2014). AVOIDIT: A cyber attack taxonomy. In *Proc. of 9th Annual Symposium On Information Assurance-ASIA* (Vol. 14). https://nsarchive.gwu.edu/sites/default/files/documents/4530310/Chris-Simmons-Charles-Ellis-Sajjan-Shiva.pdf

Song, C., Lin, F., Ba, Z., Ren, K., Zhou, C., & Xu, W. (2016). My smartphone knows what you print: Exploring smartphone-based side-channel attacks against 3d printers. *Proceedings of the 2016 ACM SIGSAC Conference on Computer and Communications Security*, 895–907. https://doi.org/https://doi.org/10.1145/2976749.2978300

Stouffer, K., Falco, J., & Scarfone, K. (2013). Guide to Industrial Control Systems (ICS) Security. In *NIST Special Publication 800-82, Revision 2*. https://doi.org/10.6028/NIST.SP.800-82r1

Stouffer, K., Stouffer, K., Zimmerman, T., Tang, C., Lubell, J., Cichonski, J., & McCarthy, J. (2017). *Cybersecurity framework manufacturing profile*. US Department of Commerce, National Institute of Standards and Technology. https://doi.org/https://doi.org/10.6028/NIST.IR.8183

*Stuxnet – Dangerous World*. (2018). https://dangerousworld.soe.ucsc.edu/2018/03/12/stuxnet/

Tanner, R., & Loh, N. K. (1992). A taxonomy of multi-sensor fusion. *Journal of Manufacturing Systems*, *11*(5), 314–325. https://doi.org/https://doi.org/10.1016/0278-6125(92)90060-S

*The Guardian*. (2012). https://www.theguardian.com/technology/2012/oct/08/china-huawei-zte-security-threat

The Ponemon Institute. (2018). *Separating the Truths from the Myths in Cybersecurity Sponsored by BMC* (Issue June). https://www.ponemon.org/local/upload/file/BMC Consolidated Report Final.pdf

*The White House*. (2021). https://www.whitehouse.gov/briefing-room/presidential-actions/2021/05/12/executive-order-on-improving-the-nations-cybersecurity/

Thornburgh, T. (2004). Social engineering: the" dark art". *Proceedings of the 1st Annual Conference on Information Security Curriculum Development*, 133–135. https://doi.org/https://doi.org/10.1145/1059524.1059554

Trippel, T., Weisse, O., Xu, W., Honeyman, P., & Fu, K. (2017). WALNUT: Waging doubt on the integrity of MEMS accelerometers with acoustic injection attacks. *2017 IEEE European Symposium on Security and Privacy (EuroS&P)*, 3–18. https://doi.org/https://doi.org/10.1109/EuroSP.2017.42

Tuptuk, N., & Hailes, S. (2018). Security of smart manufacturing systems. *Journal of Manufacturing Systems*, *47*, 93–106. https://doi.org/10.1016/j.jmsy.2018.04.007

Tweneboah-Koduah, S., Skouby, K. E., & Tadayoni, R. (2017). Cyber Security Threats to IoT Applications and Service Domains. *Wireless Personal Communications*, *95*(1), 169–185. https://doi.org/10.1007/s11277-017-4434-6

*Verizon*. (2021). https://www.verizon.com/business/resources/reports/dbir/

Wells, L. J., Camelio, J. A., Williams, C. B., & White, J. (2014). Cyber-physical security challenges in manufacturing systems. *Manufacturing Letters*, *2*(2), 74–77. https://doi.org/10.1016/j.mfglet.2014.01.005

Williams, B., Soulet, M., & Siraj, A. (2023). A Taxonomy of Cyber Attacks in Smart Manufacturing Systems. *6th EAI International Conference on Management of Manufacturing Systems*, 77–97. https://doi.org/10.1007/978-3-030-96314-9_6

Wu, D., Ren, A., Zhang, W., Fan, F., Liu, P., Fu, X., & Terpenny, J. (2018). Cybersecurity for digital manufacturing. *Journal of Manufacturing Systems*, *48*, 3–12. https://doi.org/10.1016/j.jmsy.2018.03.006

Wu, M., & Moon, Y. B. (2017). Taxonomy of Cross-Domain Attacks on CyberManufacturing System. *Procedia Computer Science*, *114*, 367–374. https://doi.org/10.1016/j.procs.2017.09.050

Wu, M., & Moon, Y. B. (2018). Taxonomy for secure cybermanufacturing systems. *ASME International Mechanical Engineering Congress and Exposition, Proceedings (IMECE)*, *2*, 1–10. https://doi.org/10.1115/IMECE2018-86091

Yampolskiy, M., King, W. E., Gatlin, J., Belikovetsky, S., Brown, A., Skjellum, A., & Elovici, Y. (2018). Security of additive manufacturing: Attack taxonomy and survey. *Additive Manufacturing*, *21*(November 2017), 431–457. https://doi.org/10.1016/j.addma.2018.03.015

Younis, Y. A., & Kifayat, K. (2013). Secure cloud computing for critical infrastructure: A survey. *The 14th Annual PostGraduate Symposium on The Convergence of Telecommunications, Networking and Broadcasting*.

Zhang, Y., Wang, L., Sun, W., Green, R. C., & Alam, M. (2011). Distributed intrusion detection system in a multi-layer network architecture of smart grids. *IEEE Transactions on Smart Grid*, *2*(4), 796–808. https://doi.org/10.1109/TSG.2011.2159818